\documentclass[%
reprint,
superscriptaddress,
nofootinbib,
 amsmath,amssymb,
 aps,
]{revtex4-2}

\usepackage{amsmath}
\usepackage{hyperref}
\usepackage{xspace}
\usepackage{newtxtext}

\usepackage{listings}
\usepackage{xcolor}
\usepackage[capitalize]{cleveref}

\makeatletter
\def\lst@outputspace{{\ifx\lst@bkgcolor\empty\color{white}\else\lst@bkgcolor\fi\lst@visiblespace}}
\makeatother

\definecolor{codegreen}{rgb}{0,0.6,0}
\definecolor{codegray}{rgb}{0.5,0.5,0.5}
\definecolor{codepurple}{rgb}{0.58,0,0.82}
\definecolor{backcolour}{rgb}{0.95,0.95,0.92}

\lstdefinestyle{mystyle}{
    backgroundcolor=\color{backcolour},   
    commentstyle=\color{codegreen},
    keywordstyle=\color{magenta},
    numberstyle=\tiny\color{codegray},
    stringstyle=\color{codepurple},
    basicstyle=\ttfamily\footnotesize,
    breakatwhitespace=false,         
    breaklines=False,                 
    captionpos=b,                    
    keepspaces=true,                 
    numbersep=5pt,                  
    showspaces=false,                
    showstringspaces=false,
    showtabs=false,                  
    tabsize=2
}

\usepackage[T1]{fontenc}

\usepackage{graphicx}
\usepackage{amsmath}
\usepackage{amssymb}

\newcommand{\apjs}{Astrophysical Journal Supplement Series}
\newcommand{\apjl}{Astrophysical Journal Letters}

\begin{document}
\title[Joint Ranking Statistic]{A Joint Ranking Statistic for Multi-messenger Astronomical Searches with Gravitational Waves}

\author{Brandon Piotrzkowski}
\email{brandon.piotrzkowski@ligo.org}
\author{Amanda Baylor}
\author{Ignacio~Maga\~{n}a Hernandez}

\affiliation{University of Wisconsin-Milwaukee, Milwaukee, WI 53201, USA}

\date{\today}

\begin{abstract}

Joint ranking statistics are used to distinguish real from random coincidences, ideally considering whether shared parameters are consistent with each other as well as whether the individual candidates are distinguishable from noise. We expand on previous works to include additional shared parameters, we use galaxy catalogues as priors for sky localization and distance, and avoid some approximations previously used. We develop methods to calculate this statistic both in low-latency using HEALPix sky maps, as well as with posterior samples. We show that these changes lead to a factor of one to two orders of magnitude improvement for GW170817-GRB 170817A depending on the method used, placing this significant event further into the foreground. We also examined the more tenuous joint candidate GBM-GW150914, which was largely penalized by these methods. Finally, we performed a simplistic simulation that argues these changes could better help distinguish between real and random coincidences in searches, although more realistic simulations are needed to confirm this.

\end{abstract}


\maketitle

\section{\label{sec:intro}Introduction}
Multi-messenger events are an important aspect of modern astronomy, potentially giving more information about an event than just individual detections. This was first shown with the joint optical and neutrino burst detection of supernova SN1987A, which helped confirm supernova models and provided upper limits on both neutrino flavors and mass \citep{hirata1987observation, arnett1987neutrino, lunardini2004neutrinos}. Astronomers have also long expected gravitational waves (GW) from binary neutron star (BNS) systems to be accompanied by prompt short gamma-ray bursts (GRB) \citep{blinnikov1984exploding, paczynski1986gamma, eichler1989nucleosynthesis}. This was also confirmed with the joint detection of GW170817 \citep{abbott2017gw170817, abbott2017gravitational} and GRB 170817A \citep{goldstein2017ordinary}. This BNS-GRB detection was the culmination of the effort of joint searches over the years \citep{aasi2014methods, singer2014first, abbott2017search, urban2016, cho2019low} and has prompted new searches to look for similar candidates in sub-threshold data \citep{hamburg2020joint, abbott2020search, Cosmin, nitz2019potential}. Identifying additional joint BNS-GRB detections would potentially help constrain the speed of gravity \citep{abbott2017gravitational}, neutron star equation of state \citep{abbott2018gw170817, coughlin2019multimessenger}, Hubble constant \citep{ligo2017gravitational}, and GRB jet models \citep{farah2020counting}.

One area of active research for joint searches has been the ranking statistic used, whose function is to rank candidates based on how likely they are to have a real joint origin rather than be randomly associated, an especially difficult task in the sub-threshold regime. The types of statistics that have been used include modified signal-to-noise ratios \citep{nitz2019potential}, false alarm rates \citep{urban2016, cho2019low}, and other generic ranking statistics \citep{hamburg2020joint}. Among this last group, \citet{Ashton} put forward a general methodology that considered the overlap of time and sky localizations as the primary determiners of significance. This seems intuitive since real coincidences must share the same underlying parameters. Next, \citet{Cosmin} improved this by considering additional noise hypotheses and including the significance of the individual candidates, albeit with some approximations. This allowed them to start to extend this method to the sub-threshold regime, where many coincidence events may still be distinguishable from noise \citep{burns2019neutron}.

In this work we expand on these methods further by considering the overlap of additional parameters (distance and inclination for BNS-GRBs) and show that filtering using coincidence models emerges naturally, shown by avoiding approximations until warranted by specific use cases. We also demonstrate the use of a galaxy catalogue as a prior for sky localization and distance in the case of GW-GRBs. Although this statistic will be inevitably more computationally expensive than previous, we derive a version with minimal additional computational cost and intended to be used in low-latency targeted searches, as well as a best-possible version using posterior samples given from parameter estimation.

This work is meant to be a Bayesian derivation of the odds ratio for whether two events are correlated (real coincidence) versus non-correlated (random coincidence), primarily based on the results of \citet{Ashton} and \citet{Cosmin} while similarly applying this method specifically to GW-GRB coincidences. We note that other Bayesian approaches have been developed for multi-messenger searches in the context of GW-neutrino bursts \citep{bartos2019bayesian, veske2020search}. We will first write out the odds ratio using Bayes rule, expand on the Bayes factors and prior odds, write out these terms in the context of GW-GRB candidates, and then conclude with a number of example demonstrations of this statistic.

\section{\label{sec:odds}Derivation}
The joint ranking statistic of interest here will be the odds ratio between whether two candidates are truly correlated (\(\mathcal{H}^c\)), meaning they have the same underlying shared parameters \(\theta\) \citep{Ashton}, or whether they are uncorrelated (\(\mathcal{H}^{\text{uncor}}\)), meaning the association is from random chance. To understand the uncorrelated hypothesis as described in \citet{Cosmin}, let us first consider a single isolated astrophysical event candidate. This candidate has only a couple of possibilities for its origin: either this event is from the intended astrophysical source (\(\mathcal{H}^s\)) or is an artifact of noise (\(\mathcal{H}^n\)), instrumental or otherwise. Next, if we consider two unrelated candidates detected by different observatories, then the joint hypothesis between them is then comprised of the two individual hypotheses. In other words, since either of the two events could be an unrelated astrophysical signal or a noise transient, this means the joint candidate could be any of the four permutations \(\mathcal{H}^{XY}\) where \(X,Y \in \{s, n\}\). Therefore the total uncorrelated hypothesis is the combination of these four possibilities, although they may be not all have the same probability and thus require different relative weights.

We can write out the odds ratio between the correlated and uncorrelated hypothesis generally as
\begin{equation}
    \mathcal{O}_{c/\text{uncor}} = \frac{P(\mathcal{H}^{c}|x_{a}, x_{b})}{P(\mathcal{H}^{\text{uncor}}|x_{a}, x_{b})} \label{odds1}
\end{equation}
where \(x_{a}\) and \(x_{b}\) are independent data sets (e.g. data from a GW and a GRB candidate respectively). Using Bayes' theorem this can be refactored to
\begin{align}
    \mathcal{O}_{c/\text{uncor}} &= \frac{P(x_{a}, x_{b} | \mathcal{H}^{c})}{P(x_{a}, x_{b} | \mathcal{H}^{\text{uncor}})} \frac{P(\mathcal{H}^{c})}{P(\mathcal{H}^{\text{uncor}})} \nonumber \\
    &= B_{c/\text{uncor}}(x_a, x_b) \frac{P(\mathcal{H}^{c})}{P(\mathcal{H}^{\text{uncor}})} \label{odds2}
\end{align}
where this first term \(B_{c/\text{uncor}}(x_a, x_b)\) is the joint Bayes factor, and the second is the prior odds. Next we will expand these two factors in terms of statistics that can be provided by astronomical experiments.

\subsection{\label{sec:bayes}General multi-messenger candidate}

Let us first work with the joint Bayes factor. Since the four components of \(\mathcal{H}_{\text{uncor}}\) are mutually exclusive of each other, we can write the total probability as the sum of each part \citep{Cosmin}. We do this by first using Bayes' theorem and writing
\begin{align}
    P(x_{a}, x_{b} &|\mathcal{H}^{\text{uncor}}) = \frac{P(\mathcal{H}^{\text{uncor}} | x_{a}, x_{b}) P(x_{a}, x_{b})}{P(\mathcal{H}^{\text{uncor}})} \\
    &= \frac{\sum_{X,Y \in \{s,n\}}P(\mathcal{H}^{XY} | x_{a}, x_{b}) P(x_{a}, x_{b})}{P(\mathcal{H}^{\text{uncor}})} ~.\label{eq:uncor}
\end{align}
Now using Bayes' theorem again, we get
\begin{equation}
    P(x_{a}, x_{b} |\mathcal{H}^{\text{uncor}}) = \frac{\sum_{X,Y \in \{s,n\}}P(x_{a}, x_{b} |\mathcal{H}^{XY}) P(\mathcal{H}^{XY})}{P(\mathcal{H}^{\text{uncor}})}
\label{likelihood_uncor}
\end{equation}
Thus the total uncorrelated likelihood can be written as the prior weighted sum of each component likelihood. Next let us examine the correlated hypothesis in \eqref{odds2}, where we can marginalize over the set of all shared parameters \(\theta\) between \(x_{a}\) and \(x_{b}\) to get
\begin{align}
    P(x_{a}, x_{b} | \mathcal{H}^{c})
    &= \int_{\Theta_c} P( x_{a}, x_{b} | \theta, \mathcal{H}^c ) P(\theta | \mathcal{H}^c) d \theta \label{likelihood_cor0} ~.
\end{align}
We note that we have restricted our domain of \(\theta\) to the subset \(\Theta_c\) where \(P(\theta | \mathcal{H}^c)>0\).
  
Let us expand the first term in the integrand as
\begin{equation}
    P( x_{a}, x_{b} | \theta, \mathcal{H}^c ) = P( x_{a}| \theta, \mathcal{H}^c ) P( x_{b} | \theta, \mathcal{H}^c ) 
\end{equation}
since \(x_{a}\) and \(x_{b}\) are independent data sets.
Now using Bayes' Theorem  we get
\begin{align}
        &~~~ P( x_{a}, x_{b} | \theta, \mathcal{H}^c ) = \notag \\
        &\frac{P( x_{a}| \mathcal{H}^c ) P(\theta | x_{a}, \mathcal{H}^c )}{P(\theta | \mathcal{H}^c )}  \frac{P( x_{b}| \mathcal{H}^c ) P(\theta | x_{b}, \mathcal{H}^c )}{P(\theta | \mathcal{H}^c )} .
\end{align}
It's worth noting that we have used \(\mathcal{H}^c\) in conjunction with data sets from a single experiment here, which still has a clear meaning since the parameter spaces for correlated and generic detections are not identical. We will expand further on such terms in section \ref{sec:bayescoinc}.
Putting the above equation back into \eqref{likelihood_cor0}, we are left with
\begin{align}
        P(x_{a}, x_{b} | \mathcal{H}^{c}) &= \notag P(x_{a} | \mathcal{H}^{c}) P(x_{b} | \mathcal{H}^{c})\\
        &\times \int_{\Theta_c} \frac{P(\theta | x_{a}, \mathcal{H}^c ) P(\theta | x_{b}, \mathcal{H}^c )}{P(\theta | \mathcal{H}^c )} d \theta
\label{likelihood_cor}
\end{align}
where we will refer to this integral as the overlap integral \(\mathcal{I}_{\theta}(x_{a}, x_{b})\), as similarly derived in \citet{Ashton}.

Since the likelihoods in \eqref{likelihood_uncor} are composed of the independent individual likelihoods of each experiment, we can write these as the product
\begin{equation}
    P(x_{a}, x_{b} | \mathcal{H}^{XY}) = P(x_{a} | \mathcal{H}_a^{X}) P(x_{b} | \mathcal{H}_b^{Y}) ~ . \label{likelihood_sep}
\end{equation}
We can then write out the overall Bayes factor in \eqref{odds2} using \eqref{likelihood_uncor} and \eqref{likelihood_cor} to get
\begin{equation}
    B_{c/\text{uncor}}(x_a, x_b) = \frac{P(\mathcal{H}^{\text{uncor}})}{\sum_{X,Y \in \{s,n\}}B_{XY/c}(x_{a}, x_{b}) P(\mathcal{H}^{XY})} \label{bayes_overall}
\end{equation}
where each individual Bayes factor, defined as the ratio of \eqref{likelihood_sep} and \eqref{likelihood_cor}, is given by
\begin{align}
    B_{XY/c} (x_a, x_b) &= \frac{P(x_{a} | \mathcal{H}_a^{X})}{P(x_{a} | \mathcal{H}^{c})} \frac{P(x_{b} | \mathcal{H}_b^{Y})}{ P(x_{b} | \mathcal{H}^{c})} / \mathcal{I}_{\theta}(x_a, x_b) \notag \\
     &= B_{X/c}(x_a) B_{Y/c}(x_b) / \mathcal{I}_{\theta}(x_a, x_b) \label{bayes_comb}
\end{align}
and where we have reduced this to the Bayes factor of each individual event along with the overlap integral between their shared parameters. Note that the noise vs coincidence Bayes factor can be separated out as
\begin{equation}
    B_{n/c}(x_a)  = B_{n/s}(x_a) B_{s/c}(x_a) . \label{eq:bayes_separated}
\end{equation}
with \(B_{n/s}(x_a)\) being the noise vs signal Bayes factor, which can be determined using the event and properties of the detector, while \(B_{s/c}(x_a)\) is the uncorrelated signal vs correlated signal Bayes factor, which can be determined by whether the measured parameters are consistent with a given coincidence model. If we plug \eqref{bayes_overall} into \eqref{odds2}, using \eqref{bayes_comb} and \eqref{eq:bayes_separated}, we get
\begin{widetext}
\begin{equation}
     \mathcal{O}_{c/\text{uncor}}(x_a, x_b) =  \frac{ P(\mathcal{H}^{c}) \mathcal{I}_{\theta}(x_a, x_b)  B_{c/s}(x_a) B_{c/s}(x_b)}{P(\mathcal{H}^{ss}) + B_{n/s}(x_a)P(\mathcal{H}^{ns}) + B_{n/s}(x_b) P(\mathcal{H}^{sn})+ B_{n/s}(x_a) B_{n/s}(x_b) P(\mathcal{H}^{nn})} \label{odds_f1}~.
\end{equation}
\end{widetext}
The terms in \eqref{odds_f1} are worth discussing in more detail. This statistic can be interpreted as having contributions from three distinct parts: 1.) the significance of the individual candidates weighted by their priors, 2.) the overlap of the shared parameters between the candidates, and 3.) the evidence for whether the data for each candidate favors a coincident model rather than a generic detection model. We see in the limit where each candidate is infinitely significant, there is still the possibility that they are randomly associated (i.e. \(\mathcal{H}^{ss}\)), leaving the overlap integral \(\mathcal{I}_{\theta}(x_a, x_b)\) and coincidence Bayes factors as the sole determiners of joint significance. Previous methods have have not considered these coincidence Bayes factors and have set \(B_{c/s}(x_a)=1\) \citep{Ashton, Cosmin}. However this is not strictly correct, as there are regions of parameter space with different expectations of whether a coincidence is possible. We will discuss these specific terms using a specific example in section \ref{sec:bayescoinc}.

\subsection{\label{sec:prior}GW-GRB candidate}

We will next focus on writing down the various terms in \eqref{odds_f1} for the specific case of GW-GRB coincidences, starting with the prior terms. The probability of detecting a single event should follow a Poisson distribution \citep{urban2016, Ashton} so that
\begin{equation}
    P( \mathcal{H}^{X} ) =  R_X T  e^{- R_X T } .
\end{equation}
where \(T\) is the co-observing time and \(R_x\) is the rate. If we consider the co-observing time \(T\) for an individual candidate as the co-observing time window \(\Delta t\), typically at least two orders of magnitude higher than the corresponding rate, we can take the approximation that
\begin{equation}
    P( \mathcal{H}^{X} ) \approx  R_X T .
\end{equation}
For the case of two independent events, the probability that they both occur is the product
\begin{equation}
    P( \mathcal{H}^{XY} ) \approx R_X R_Y T^2 .
\end{equation}
Meanwhile the noise vs signal Bayes factor \(B_{n/s}(x_a)\) can be supplied by the given experiment or calculated by using a proxy detection statistic once a number of events have been collected \citep{Cosmin}. Another example of this is calculated by \texttt{ligo.skymap}, 
although this assumes Gaussian noise rather than the non-stationary noise artifacts typically found in LVK data.

The overlap integral \(\mathcal{I}_{\theta}(x_a, x_b)\) derived in \eqref{likelihood_cor} measures the overlap between the common set of parameters \(\theta\) of two experiments. For joint GW-GRB detections (under the $\mathcal{H}^c$ hypothesis) this is
\begin{align}
    &\mathcal{I}_{\theta}(x_{gw}, x_{grb}) = \int_{\Theta_c} \frac{P(\theta | x_{gw}, \mathcal{H}^c ) P(\theta | x_{grb}, \mathcal{H}^c )}{P(\theta | \mathcal{H}^c)} d \theta\\
    &~~~= \int_{\Theta_c} \ d D_L  d \iota d \Omega d t_{c} \ \frac{P(D_L, \iota, \Omega, t_{c} | x_{gw}, \mathcal{H}^c )}{P(D_L, \iota, \Omega, t_{c} | \mathcal{H}^c)} \notag \\
    &~~~~~~~~~~~~\times P(z(D_L), \theta_v(\iota), \Omega, t_{c} + t_d | x_{grb}, \mathcal{H}^c) \label{eq:overlap_analytical}
\end{align}
where the shared parameters are $\theta = \{ D_L, \iota, \Omega, t_{c}\}$. More explicitly, $D_L$ is the luminosity distance, $\iota$ the inclination angle, $\Omega$ the sky location of the source and $t_{c}$ the coalescence time. We note that the overlap in distance has been used before in the case of GW190521-ZTF19abanrhr \citep{ashton2020current}. The parameter \(t_d\) is the time delay between the the GW and GRB arrival times, which we allow to span the possible range of \(t_l\) before to \(t_h\) after the coalescence time, meaning the total co-observing time window is \(\Delta t =| t_h - t_l |\). If the observations are disjoint, with inconsistent measurements of the parameters, the overlap integral above will be negligible and thus will heavily weight against the correlated hypothesis. From the generic overlap integral \(\mathcal{I}_{\theta}\) we can separate out the overlap in time 
\begin{align*}
\mathcal{I}_{t_c} =
\begin{cases}
     T / \Delta t &(t_l < t_{d} < t_h) \\
     0 &(otherwise)
\end{cases}
\end{align*}
as in \citet{Ashton} and \citet{Cosmin} to write \eqref{odds_f1} as
\begin{widetext}
\begin{equation}
     \mathcal{O}_{c/\text{uncor}}(x_{gw}, x_{grb}) = \frac{R_{gw,grb}^c \mathcal{I}_{\Omega, D_L,\iota}(x_{gw},x_{grb}) B_{c/s}(x_{gw}) B_{c/s}(x_{grb}) }{\Delta t\left[R_{gw}^s R_{grb}^s + B_{n/s}(x_{gw}) R_{gw}^n R_{grb}^s + B_{n/s}(x_{grb}) R_{gw}^s R_{grb}^n+ B_{n/s}(x_{gw}) B_{n/s}(x_{grb}) R_{gw}^n R_{grb}^n \right]}   \label{odds_f2}
\end{equation}
\end{widetext}
where the \(R_a^n\) terms are the rates of noise triggers that are considered in a given joint search, the \(R_a^s\) terms are expected rates of detected real astrophysical triggers, and \(R_{a,b}^c\) is the expected rate of detected real coincidences. We note that \eqref{odds_f2} in principle should perform well in sub-threshold searches because for a joint candidate to have a higher significance, both individual candidates need a higher individual significance. This effect will be more pronounced than in \citet{Cosmin} due to the inclusion of rates, helping to penalize obvious noise events and potentially bring real coincidences to the foreground.

We consider \eqref{odds_f2} the primary result of this paper and later detail how to calculate this for specific examples in section \ref{sec:example}.

\subsection{\label{sec:bayescoinc}Coincidence Bayes Factors}

The presence of coincident vs signal Bayes factors \(B_{c/s}(x_a)\) in \eqref{odds_f2} is one way that differentiates this statistic from previous efforts. This was claimed in  \citep{cho2019low} to always be unity, but ignored the differences in priors between coincidences and generic detections. We can see this by expanding over the parameters \(\theta_a\)
\begin{equation}
    B_{c/s}(x_a) = \frac{\int P(x_a | \theta_a, \mathcal{H}^c) P (\theta_a| \mathcal{H}^c) d \theta_a}{\int P(x_a | \theta_a, \mathcal{H}_a^s) P (\theta_a| \mathcal{H}_a^s) d \theta_a}  \label{eq:bayes_coinc}
\end{equation}
using the likelihood \(P(x_a | \theta_a, \mathcal{H}_a^s)\) and prior \( P (\theta_a| \mathcal{H}_a^s)\).
The coincidence Bayes factor should be interpreted has how much the measured parameters of an \textit{individual} experiment supports a coincidence by comparing between coincidence and generic detection models. For example, if the masses of a GW candidate are both above \(5 M_\odot\) we would think the progenitor is a binary black hole system and not capable of producing a GRB, hence we could set \(B_{c/s}(x_{gw}) = 0\). For GRBs, an example would be placing a cutoff on the duration and require \(T_{90} \leq 2.0 s\) as to only consider short GRBs. This is already possible to include in multi-messenger searches by simply screening out candidates that don't meet certain parameter requirements. In other words, one could define \(B_{c/s}(x_{a})\) as a convolution of step functions by giving \(1\) if the parameters fall within expected bounds and \(0\) if not, an especially useful method for a low-latency search. However, a more explicit calculation could be used to gain more sensitivity and we will show such an example by using posterior distributions in the next section.

\section{\label{sec:example}Examples}

Let us test our method by applying \eqref{odds_f2} to a number of situations, including a couple of known GW-GRB coincident candidates, GW170817-GRB 170817A and GBM-GW150914, as well as a simulation of many coincidences.
If the individual events are significant enough so that the noise-to-signal Bayes factors \(B_{n/s}(x_a)\approx 0\), then our odds ratio from \eqref{odds_f2} simplifies to
\begin{align}
     &\mathcal{O}_{c/\text{uncor}}(x_{gw}, x_{grb}) \approx \notag \\ &~~~\frac{R_{gw,grb}^c}{R_{gw}^s R_{grb}^s \Delta t }
    \mathcal{I}_{\Omega ,D_L,\iota} B_{c/s}(x_{gw}) B_{c/s}(x_{grb}) \label{odds_f3} ,
\end{align}
where this result is similar to that in \citet{Cosmin} with the addition of the coincidence Bayes factors \(B_{c/s}(x_{grb})\) and overlap \(\mathcal{I}_{\Omega,D_L,\iota}\) with distance and inclination. Let us briefly detail the various fixed values in \eqref{odds_f3} in the case of GW170817-GRB 170817A. The measured expected rate of BNS triggers in O2 was \(R_{gw}^s\approx 0.8/\text{year}\) while the expected rate of \textit{significant} GW-GRB coincidences was \(R_{gw,grb}^c \approx  0.14/\text{year}\) \citep{howell2019joint}. We also take the detected rate of Fermi-GBM short GRBs to be \(R_{grb}^s \approx 40/\text{year}\) \citep{howell2019joint}, although including additional GRB experiments would increase this. The coincidence window we consider is the standard GW-GRB \([-1, +5] s\) centered on the merger time, which means \(\Delta t = 6 s\) \citep{abbott2017search}. We also make the approximation \(P(x_{a} | \theta_{a}, \mathcal{H}^c) \approx P(x_{a} | \theta_{a}, \mathcal{H}_a^s)\) since coincident parameter estimation may not be available for many events (these in practice usually just assume a fixed sky position) and sufficiently interesting joint candidates should already have support in coincident parameter regimes. This also gives terms like \eqref{eq:bayes_coinc} a clear interpretation: the coincident hypothesis has more support if the likelihood favors the more constrained coincident prior.

\subsection{\label{sec:lowlatency}GW170817-GRB 170817A: Using 3D sky maps}

In order for this statistic to be useful in a low-latency targeted search, it must only use information immediately available to these searches and itself be calculable in low-latency. To this end we restricted ourselves to only using HEALPix sky maps, with the GW sky map including distance information \citep{HEALPix,singer2016rapid}. We also used the GLADE v2.4 galaxy catalogue as a prior over sky-localization and distance, both with the motivation of being more astrophysically motivated and also to speed up the calculation, although the latter depends on the sky area for a particular joint event \citep{dalya2018glade}. While loading even partial catalogue may take some time, to the point of possibly requiring to load this prior, we find evaluating on these galaxies is much faster. We have ignored inclination in this case since this information isn't currently available to these searches using BAYESTAR, but requires posterior samples produced from much higher latency.

We can write the joint sky-localization/distance overlap integral \eqref{eq:overlap_analytical} as
\begin{align}
    \mathcal{I}_{\Omega, D_L} &= \notag \int d \Omega d D_L \frac{p(\Omega, D_L | x_{gw}, \mathcal{H}_{gw}^s)}{p(\Omega, D_L | \mathcal{H}_{gw}^s)}  \\ &~~ \times\frac{p(\Omega | x_{grb}, \mathcal{H}_{grb}^s)}{p(\Omega | \mathcal{H}_{grb}^s)} p(\Omega, D_L | \mathcal{H}^c) \label{eq:overlap_lowlatency_start}
\end{align}
where we assume the GRB has no distance information and let \(p(\Omega, D_L | \mathcal{H}^c)\) be the galaxy catalogue prior, written as
\begin{align}
    p(\Omega, D_L | \mathcal{H}^c) &= \sum_{j=0}^{N_{gal}} w_j \delta(\Omega - \Omega_j) \delta(D_L - D_{L,j}) ~ . 
\end{align}
Here \(w_j\) are the weights to the \(j\)-th galaxy and \(N_{gal}\) is the total number of galaxies in the catalogue within the priors. Note if we used uniform weights then \(w_j \propto 1/N_{gal}\), or we could weight each galaxy by its luminosity so that \(w_j \propto L_j / \sum_k L_k\). If we assume the priors on individual localizations are uniform, we convert from sky coordinates to pixel space, and then normalize the galaxy catalogue prior we get
\begin{align}
    \mathcal{I}_{\Omega, D_L} &= \notag  N_{pix}^2 \sum_{j=0}^{N_{gal}} w_j  P(\Omega_j | x_{gw}, \mathcal{H}_{gw}^s) P(\Omega_j | x_{grb}, \mathcal{H}_{grb}^s)\\
    &~~\times\frac{p( D_{L,j} | \Omega_j, x_{gw}, \mathcal{H}_{gw}^s)}{p( D_{L,j} | \mathcal{H}_{gw}^s)} \label{eq:overlap-healpix}
\end{align}
where \(N_{pix}\) is the number of pixels in an individual sky map and \(P(\Omega_j | x_{a}, \mathcal{H}_{a}^s)\) is the probability in a given HEALPix pixel.

We used a \(D_L^2\) prior from \(10~\text{Mpc}\) to \(60~\text{Mpc}\) and calculated the line of sight distance posterior using the conditional posteriors in \texttt{ligo.skymap} \citep{singer2016supplement}. We stress that the choice of the distance prior needs to follow the probability distribution fairly tightly, as setting the maximum distance too large will artificially increase the overlap in distance. One way to test that this was done properly is ensuring that \(\mathcal{I}_{\Omega} \approx \mathcal{I}_{\Omega, D_L}\) using a uniform catalogue, where each pixel contains one galaxy and the distance is sampled from the \(D_L^2\) prior. We also explored the utility of making distance and luminosity cuts, as well as weighting each galaxy by its luminosity \citep{fishbach2019standard}, which we found improved our measurement.

While the inclusion of the overlap in distance may seem strange since we assume the GRB will not have a distance measurement, we offer another explanation. Considering now just the terms including distance in \eqref{eq:overlap_lowlatency_start}, we see that these are measuring whether the posterior over distance supports the galaxy catalogue vs the general \(D_L^2\) prior. This difference might be quite significant when the localization is small, as is the case with GW170817, and we see support of the catalogue in tables \ref{tab:results_lowlatency} and \ref{tab:results_samples}. Also in the situations where there is a distance measurement from afterglow measurements or else, as may be in the case of a detection by \textit{Swift} \citep{gehrels2004swift}, this could be included and likely would improve this calculation.

The results of this method are summarized in table \ref{tab:results_lowlatency}. If we include galaxies from the whole GLADE catalogue after distance and luminosity cuts, we get \(\mathcal{I}_{\Omega, D_L}\approx 284\). This is a notable improvement over using a non-informative (uniform) galaxy catalogue (\(\mathcal{I}_{\Omega, D_L}\approx 36.9\)), where each pixel contains one galaxy and the distance is sampled from the prior. We should note that the non-informative case is numerically consistent with and analytically simplifies to the spatial overlap integral used in the RAVEN pipeline \citep{urban2016, cho2019low} and separately calculated in \citet{Ashton}.  If we only include NGC 4993 in our catalogue, the galaxy that GW170817 was localized to  \citep{valenti2017discovery, abbott2017multi}, we find that \(\mathcal{I}_{\Omega, D_L}\approx 380,000\). An interesting double check and alternate application of this method is to rank galaxies as hosts using \eqref{eq:overlap-healpix}. If we only consider sky-localization, we find that the host galaxy NGC 4993 is ranked 5th in the catalogue, while including distance information changes this rank to 3rd.

Using \eqref{odds_f3} we can directly compare with the results of \citet{Ashton}, albeit using updated rates and sky maps. We will ignore the coincident Bayes factors \(B_{c/s}(x_{a})\) momentarily as is in \citet{Ashton}. We find we get \(\mathcal{O}_{\text{c/uncor}}(x_{gw}, x_{grb})\approx 6,500,000\) compared to the previous method \(\mathcal{O}_{\text{c/uncor}}(x_{gw}, x_{grb})\approx 870,000\), an improvement of a factor of about 7.5 due to the inclusion of distance and the galaxy catalogue prior.

\subsection{\label{sec:posterior}GW170817-GRB 170817A: Using posterior samples}

We can also calculate \eqref{odds_f3} using posterior samples to see how well our statistic performs using the best available information. We used the low-spin posterior samples for GW170817 from \citet{abbott2019properties} since these samples assumed a uniform sky prior. However, posterior samples aren't typically produced for GRBs, so we were limited to using the sky localization and instead used models for the remaining shared parameters. Since inclination information is available from GW samples, we can go through a similar derivation to \eqref{eq:overlap-healpix} again with this additional parameter to find
\begin{align}
    &\mathcal{I}_{\Omega, D_L, \iota} =N_{pix}^2 \sum_{j=0}^{N_{gal}} w_j P(\Omega_j | x_{gw}, \mathcal{H}_{gw}^s) P(\Omega_j | x_{grb}, \mathcal{H}_{grb}^s)  \notag \\ ~~~&\times\int d \iota \frac{p( D_{L,j}, \iota | \Omega_j, x_{gw}, \mathcal{H}_{gw}^s)}{p( D_{L,j} | \mathcal{H}_{gw}^s)} \frac{p(\iota | D_{L,j}, x_{grb}, \mathcal{H}_{grb}^s)}{p(\iota | \mathcal{H}_{gw}^s)} \label{eq:overlap-healpix-samples}.
\end{align}
We took \(p(\iota | \mathcal{H}_{gw}^s)\) as a uniform prior over $\sin \iota $ ranging from 0 to \(\pi\) while we modeled \(p(\iota | D_{L,j}, x_{grb}, \mathcal{H}_{grb}^s)\) similarly except now constrained by the maximum viewing angle given by the distance, as done in \citep{howell2019joint}. This relationship between maximum viewing angle and distance is both dependent on the GRB jet model and on the properties of the the specific GRB experiment, and we note that this is somewhat circular as the parameters from the GRB jet model used in \citet{howell2019joint} are constrained themselves by GW170817-GRB 170817A. In the future, jet models will be improved and made more robust with additional GW-GRB joint detections \citep{farah2020counting}. To check robustness against this jet model, we also calculated the overlap using a constant maximum viewing angle of \(25^\circ\) and found generally similar results. The conditional GW distance/inclination posterior was calculated using a 4-dimensional kernel density estimation (KDE), checked to match the original posterior samples using resampling. The GW HEALPix sky map was calculated from the samples using \texttt{ligo.skymap}.

\begin{table}
    \centering
    \begin{tabular}{c|r|r}
         Galaxy Catalogue & $\mathcal{I}_{\Omega}$ & $\mathcal{I}_{\Omega, D_L}$\\ \hline
         Uniform                    & 37.7    & 36.9 \\
         GLADE (un-weighted)        & 152     & 217 \\
         GLADE (\(L_B>.05 L_{B*}\)) & 197     & 284 \\
         Only NGC 4993              & 210,000 & 380,000 \\
    \end{tabular}
    \caption{Overlap integrals for GW170817-GRB 170817A using the low-latency method from \eqref{eq:overlap-healpix}. We see that including distance, with the exception of uninformative uniform catalogue, improves significance due to galaxy clustering around areas of higher probability. Also including luminosity cuts of GLADE where \(L_B>.05 L_{B*}\), which gives us near \(100 \%\) completeness at \(60\) Mpc \citep{fishbach2019standard}, in addition to weighting based on \(B\)-band luminosity improves this further.}
    \label{tab:results_lowlatency}
\end{table}

\begin{table}
    \centering
    \begin{tabular}{c|r|r|r}
         Galaxy Catalogue & $\mathcal{I}_{\Omega}$ & $\mathcal{I}_{\Omega, D_L}$ & $\mathcal{I}_{\Omega, D_L, \iota}$\\ \hline
         Uniform                    & 37.7    & 35.9    & 113 \\
         GLADE (un-weighted)        & 152     & 248     & 671 \\
         GLADE (\(L_B>.05 L_{B*}\)) & 197     & 322     & 1080 \\
         Only NGC 4993              & 210,000 & 478,000 & 1,190,000 \\
    \end{tabular}
    \caption{Overlap integrals for GW170817-GRB 170817A using the posterior sample method from \eqref{eq:overlap-healpix-samples}. As in table \ref{tab:results_lowlatency}, including distance and using the GLADE catalogue both increases significance. Also, the inclusion of inclination improves the overlap further due to the samples favoring a higher inclination.}
    \label{tab:results_samples}
\end{table}

Since \(\iota\) is not directly measured by GRB experiments we could instead calculate its contribution using the GW coincident Bayes factor \(B_{c/s}(x_{gw})\) by expanding over \(\iota\). This should be an equivalent approach because both the overlap integral and coincidence Bayes factor enter \eqref{odds_f3} similarly. We obtained the likelihood in equation \eqref{eq:bayes_coinc} by re-weighting the posterior samples using the inverse of the inclination prior as weights. Similar to what is done with the overlap integral, we used a $\sin(\iota)$ prior bounded to less than the maximum viewing angle \(25^\circ\). This gave us the result of \(B_{c/s}(x_{gw})=2.7\) for the GW coincident Bayes factor. From table \ref{tab:results_samples} we can see that this is roughly the factor between \(\mathcal{I}_{\Omega, D_L}\) and \(\mathcal{I}_{\Omega, D_L, \iota}\), so we should be fairly indifferent where we include this contribution. In principle a similar calculation could also be done for other parameters such as mass using a combined BNS, binary black holes (BBH), and neutron star-black hole (NSBH) mass population model.

We also calculated a contribution from the GRB coincident Bayes factor \(B_{c/s}(x_{grb})\) by expanding over the duration \(T_{90}\), which for GRB 170817A was measured at \(T_{90}= 2.0 \pm 0.5 s\) \citep{goldstein2017ordinary}. We estimated the likelihood as a Gaussian with these values. We used a KDE over \(T_{90}\) values from Fermi on HEASARC for the general prior \citep{von2020fourth} and restricted these to \(T_{90} < 2.5 s\) events for the coincident prior, giving us \(B_{c/s}(x_{grb})=12.9\). As in the case of GWs, doing similar calculations for other parameters could improve this statistic further.

Using this method with posterior samples we find that \(\mathcal{O}_{\text{c/uncor}}(x_{gw}, x_{grb})\approx 25,000,000\) if we neglect the coincident Bayes factors, while including these gives \(\mathcal{O}_{\text{c/uncor}}(x_{gw}, x_{grb})\approx 320,000,000\), leading to an improvement of a factor of \(29\) and \(370\) respectively. This implies candidates might be better stratified, as the highly significant joint event GW170817-GRB 170817A is pushed much further out into the foreground compared to the more abundant uncorrelated joint candidates that will typically fail to have overlap in their parameters.

\subsection{GBM-GW150914 \label{sec:GW150914}}

In addition to the very confident joint detection of GW170817-GRB 170817A, we also wanted to test with a more tenuous association in order to find the limitations of this method. We therefore computed the significance of the first binary black hole (BBH) GW detection GW150914 with the corresponding temporally coincident Fermi-GBM GRB candidate \citep{connaughton2016fermi}. The astrophysical nature of this GRB has been called into question \citep{greiner2016fermi}, so we will be wary as we make an evaluation of this joint event using similar methods as in section \ref{sec:posterior}. We used the GW150914 posterior samples given in the O1 data release \citep{Abbott_2016} and the GRB candidate sky map produced by the Fermi-GBM team \citep{connaughton2016fermi}.

Unlike in section \ref{sec:posterior}, we should be cautious about the use of a galaxy catalogue as a prior due to the amount of incompleteness at these distances. Nonetheless we used a number of luminosity cuts and weights to see how this method fairs, detailed in table \ref{tab:results_gw150914}. We can see that in general this coincidence is less significant than GW170817-GRB 170817A, which is expected given the general lack of confidence in comparison. We also see that including distance and inclination information generally improves confidence by the same factors as in table \ref{tab:results_samples}, implying this information may be shared generally among significant gravitational wave candidates and could possibly not be helpful in distinguishing between joint candidates, although more investigation is needed to confirm this.

In conclusion for this event, we note that despite some non-negligible overlap in the parameters, our statistic should return \(\mathcal{O}_{\text{c/uncor}}(x_{gw}, x_{grb})\approx 0\) here. This is because there is little expectation of a significant joint BBH-GRB detection rate and we could set \(B_{c/s}(x_{gw})=0\) just due to the measured masses of GW150914, even before including the lower significance of the GRB candidate \citep{greiner2016fermi}.


\begin{table}
    \centering
    \begin{tabular}{c|r|r|r}
         Galaxy Catalogue & $\mathcal{I}_{\Omega}$ & $\mathcal{I}_{\Omega, D_L}$ & $\mathcal{I}_{\Omega, D_L, \iota}$\\ \hline
         Uniform                  & 7.07 & 7.03 & 36.5 \\
         GLADE (un-weighted)      & 5.13 & 8.02 & 25.7 \\
         GLADE (weighted)         & 3.58 & 5.52 & 20.9 \\
         GLADE (\(L_B>2 L_{B*}\)) & 2.98 & 4.49 & 18.5 \\
    \end{tabular}
    \caption{Overlap integrals for GBM-GW150914 using the posterior sample method similar to table \ref{tab:results_samples}. In general we see that increasing catalogue information decreases the significance of this coincidence in contrast to GW170817-GRB 170817A, a potential additional argument this may not be a real coincidence. We also used a \(L_B>.05 L_{B*}\) cut, as was done in table \ref{tab:results_samples}, but this returned identical values to the weighted full GLADE catalogue due to only to very close galaxies being cut, which had no support from the posterior samples. We note that none of the cuts here give \(100\%\) completeness, but in the case of \(L_B>2 L_{B*}\) we wanted to show the effects of an extreme cut.}
    \label{tab:results_gw150914}
\end{table}

\subsection{Simulation \label{sec:simulation}}

Lastly, we performed a simplistic simulation of a joint GW-GRB search to quantify the potential improvements compared to previous methods. We modeled the Bayes factor \(B_{s/n}(x_{gw})\) as Gaussian distributions in log space with means of \(1\) and \(10\) for the random and real sets respectively, and standard deviations of \(10^1\) and \(10^{1.5}\) respectively. For the other Bayes factor \(B_{s/n}(x_{grb})\), we instead used power laws of of \(-2\) and \(-1\) for the random and real sets respectively, with a lower bound at zero. Finally, we modeled the sky map overlap integral \(\mathcal{I}_{\Omega}\) as Gaussians in log space with means \(10^{-1}\) and \(10\) for the random and real sets respectively, and both with standard deviations of \(10\). These were chosen to roughly match the outputs expected from pipelines based on our preliminary investigations of doing a more robust simulation. We also used the rates detailed earlier in section \ref{sec:example}. We didn't use many of the additional terms introduced in this work, such as the additional overlap integrals, the galaxy catalogue prior, or the coincidence Bayes factors; the only difference between the statistics tested was the rate-weighted denominator in equation \eqref{odds_f2}, compared to the weightless detection Bayes factors \citet{Cosmin} and unity \citet{Ashton}. This was done in order to show how this simple change could greatly increase sensitivity.

\begin{figure}
    \centering
    \includegraphics[scale=.60]{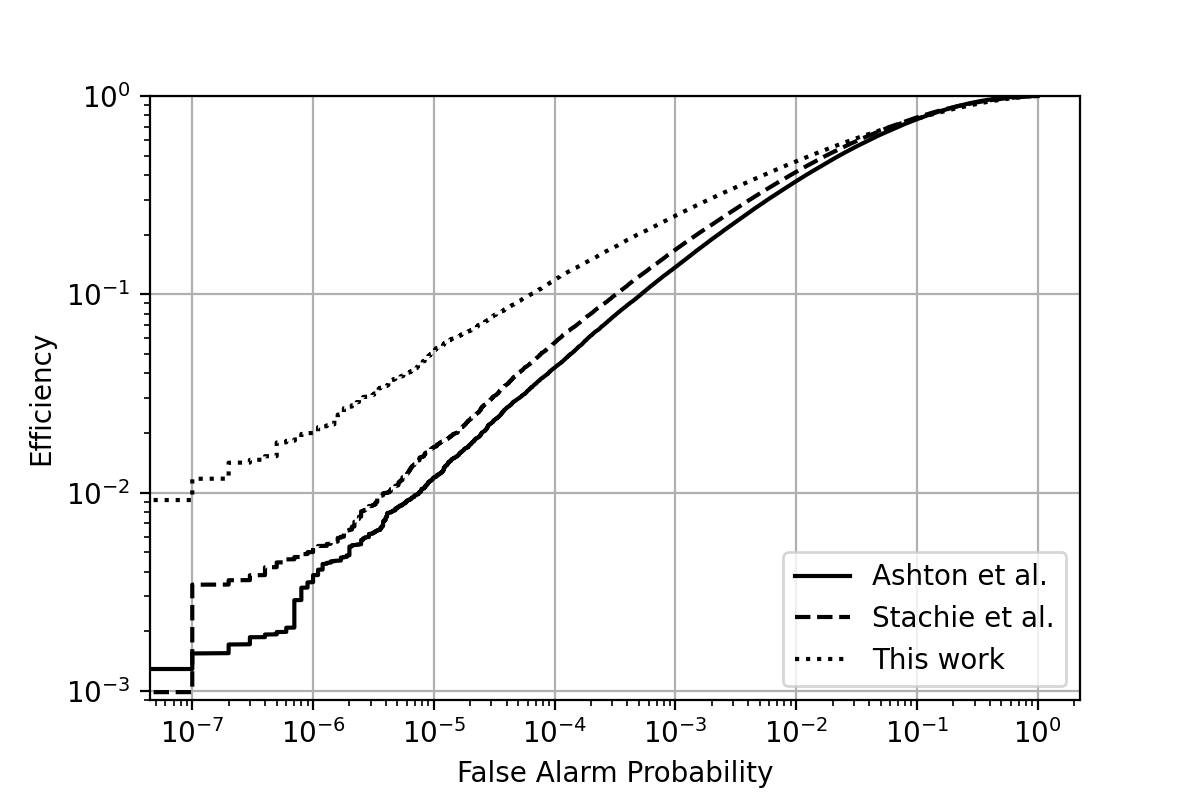}
    \caption{Results for the simulations described in section \ref{sec:simulation}, using the statistics from \citet{Ashton}, \citet{Cosmin}, and equation \eqref{odds_f2} in this work. The horizontal axis is the fraction of random coincidences above a given statistic value while the vertical axis is the fraction of real coincidences above this value. The statistics here vary only whether or not they include the significance from individual candidates, weighted based on rates in the case of equation \eqref{odds_f2}. We see generally that given a false alarm probability, our statistic returns more real coincidences.} 
    \label{fig:ROC_curve}
\end{figure}

We can see the results of this simulation in figure \ref{fig:ROC_curve}. At most given false alarm probabilities, equation \eqref{odds_f2} recovers more real injections compared to previous methods. We also tried a number of reasonable parameters for our injected distributions and this trend tended to hold. This likely occurs because \eqref{odds_f2} tends to downweight random coincidences compared to previous because of this stricter condition of requiring individual candidates to be more significant. In general, we found that \eqref{odds_f2} outperformed the others in cases where the individual detection Bayes factors distributions were more distinct and performed worse when these Bayes factors were not informative, although we argue the former case is more realistic since random coincidences are dominated by lower significance noise events (\(\mathcal{H}^{nn}\)). We should be a bit wary of these results since this simulation is rather simplistic, and we emphasize that more robust simulations are required to definitively establish that this statistic should be preferred to previous iterations.

\section{\label{sec:conclusion}Conclusion}

We derived a joint ranking statistic in a similar manner to \citet{Ashton} and \citet{Cosmin} while adding additional parameters in the overlap integral, using a galaxy catalogue as a prior, and including the previously neglected coincidence-vs-signal Bayes factors, while presenting some evidence that this statistic may perform better than these previous iterations. This statistic placed GW170817-GRB 170817A farther into the foreground by a factor of \(7.5-370\), based on the method used. We showed that this statistic minimized the significance of the more dubious association of GBM-GW15091, while also present some possible short-comings in this case. We performed a basic simulation that argues in favor of this statistic, although an additional study simulating real and background joint events is needed to confirm this and to determine which additional terms will be most useful for searches. 


We acknowledge that some of the new terms introduced with equation \eqref{odds_f2} may not be helpful in distinguishing between real and random coincidences in real searches. We showed in the case of GRB-GW150914 that including distance and inclination information improved the joint significance similarly to GW170817-GRB 170817A, meaning that this improvement may be shared by significant GW detections.
We also note that results from the coincident Bayes factors would be shared with other reasonable joint candidates, \(B_{c/s}(x_{grb})\) with any other short GRB, and \(B_{c/s}(x_{gw})\) with other close face-on candidates, so the addition of these could be ignored and replaced by filter in an actual search such as in \citet{Cosmin}. If an EM redshift is available, as may be in the case of \text{Swift} XRT follow-up \citep{gehrels2004swift}, there will likely be an improvement in the distance overlap. However in the case of other experiments such as Fermi-GBM \citep{meegan2009fermi}, the approach detailed in this work is likely the best we can do at the moment. We also note that more care could be given to distinguishing between the coincidence and generic detection terms, as well as towards more informative priors (e.g. using antenna factors in the sky localization). Finally, although there are estimates for the coincident rates in our priors, there are significant uncertainties due the poorly constrained BNS rate \citep{abbott2017gravitational, howell2019joint} and GRB jet models \citep{farah2020counting}. We argue that while including detection rates improves significance as in section \ref{sec:simulation}, the exact numbers that should be used are rather uncertain until more observations are done and instead could be replaced with the expected ratio of rates.

\section{\label{sec:acknowledgements}Acknowledgements}

We thank Tito Dal Canton, Eric Burns, Aaron Tohuvavohu, and Peter Shawhan for useful comments and discussions which improved this work.
This work makes use of the
\texttt{scipy} \citep{scipy:2020},
\texttt{numpy} \citep{oliphant2006guide, van2011numpy, harris2020array},
\texttt{astropy} \citep{robitaille2013astropy, price2018astropy},
\texttt{ligo.skymap} \footnote{\url{https://lscsoft.docs.ligo.org/ligo.skymap/}},
and \texttt{bilby} \citep{bilby} scientific software packages.
We used the low-spin posterior samples of GW170817 from \citet{abbott2019properties} \footnote{\url{https://dcc.ligo.org/LIGO-P1800061/public}} and the posterior samples for GW150914 from \citep{Abbott_2016} \footnote{\url{https://dcc.ligo.org/LIGO-T1800235/public}}.

This work was supported by NSF awards PHY-1607585 and PHY-1912649. The authors are grateful for computational resources provided by the LIGO Laboratory and supported by National Science Foundation Grants PHY-0757058 and PHY-0823459, and those provided by the Leonard E Parker Center for Gravitation, Cosmology and Astrophysics at the University of Wisconsin-Milwaukee. IMH is supported by the NSF Graduate Research Fellowship Program under grant DGE-17247915. This article has been assigned LIGO document number LIGO-P1900362.


\bibliography{references}

\providecommand{\noopsort}[1]{}\providecommand{\singleletter}[1]{#1}%
\begin{thebibliography}{49}%
\makeatletter
\providecommand \@ifxundefined [1]{%
 \@ifx{#1\undefined}
}%
\providecommand \@ifnum [1]{%
 \ifnum #1\expandafter \@firstoftwo
 \else \expandafter \@secondoftwo
 \fi
}%
\providecommand \@ifx [1]{%
 \ifx #1\expandafter \@firstoftwo
 \else \expandafter \@secondoftwo
 \fi
}%
\providecommand \natexlab [1]{#1}%
\providecommand \enquote  [1]{``#1''}%
\providecommand \bibnamefont  [1]{#1}%
\providecommand \bibfnamefont [1]{#1}%
\providecommand \citenamefont [1]{#1}%
\providecommand \href@noop [0]{\@secondoftwo}%
\providecommand \href [0]{\begingroup \@sanitize@url \@href}%
\providecommand \@href[1]{\@@startlink{#1}\@@href}%
\providecommand \@@href[1]{\endgroup#1\@@endlink}%
\providecommand \@sanitize@url [0]{\catcode `\\12\catcode `\$12\catcode
  `\&12\catcode `\#12\catcode `\^12\catcode `\_12\catcode `\%12\relax}%
\providecommand \@@startlink[1]{}%
\providecommand \@@endlink[0]{}%
\providecommand \url  [0]{\begingroup\@sanitize@url \@url }%
\providecommand \@url [1]{\endgroup\@href {#1}{\urlprefix }}%
\providecommand \urlprefix  [0]{URL }%
\providecommand \Eprint [0]{\href }%
\providecommand \doibase [0]{https://doi.org/}%
\providecommand \selectlanguage [0]{\@gobble}%
\providecommand \bibinfo  [0]{\@secondoftwo}%
\providecommand \bibfield  [0]{\@secondoftwo}%
\providecommand \translation [1]{[#1]}%
\providecommand \BibitemOpen [0]{}%
\providecommand \bibitemStop [0]{}%
\providecommand \bibitemNoStop [0]{.\EOS\space}%
\providecommand \EOS [0]{\spacefactor3000\relax}%
\providecommand \BibitemShut  [1]{\csname bibitem#1\endcsname}%
\let\auto@bib@innerbib\@empty
\bibitem [{\citenamefont {Hirata}\ \emph {et~al.}(1987)\citenamefont {Hirata},
  \citenamefont {Kajita}, \citenamefont {Koshiba}, \citenamefont {Nakahata},
  \citenamefont {Oyama}, \citenamefont {Sato}, \citenamefont {Suzuki},
  \citenamefont {Takita}, \citenamefont {Totsuka}, \citenamefont {Kifune} \emph
  {et~al.}}]{hirata1987observation}%
  \BibitemOpen
  \bibfield  {author} {\bibinfo {author} {\bibfnamefont {K.}~\bibnamefont
  {Hirata}}, \bibinfo {author} {\bibfnamefont {T.}~\bibnamefont {Kajita}},
  \bibinfo {author} {\bibfnamefont {M.}~\bibnamefont {Koshiba}}, \bibinfo
  {author} {\bibfnamefont {M.}~\bibnamefont {Nakahata}}, \bibinfo {author}
  {\bibfnamefont {Y.}~\bibnamefont {Oyama}}, \bibinfo {author} {\bibfnamefont
  {N.}~\bibnamefont {Sato}}, \bibinfo {author} {\bibfnamefont {A.}~\bibnamefont
  {Suzuki}}, \bibinfo {author} {\bibfnamefont {M.}~\bibnamefont {Takita}},
  \bibinfo {author} {\bibfnamefont {Y.}~\bibnamefont {Totsuka}}, \bibinfo
  {author} {\bibfnamefont {T.}~\bibnamefont {Kifune}}, \emph {et~al.},\
  }\bibfield  {title} {\bibinfo {title} {Observation of a neutrino burst from
  the supernova sn1987a},\ }\href@noop {} {\bibfield  {journal} {\bibinfo
  {journal} {Physical Review Letters}\ }\textbf {\bibinfo {volume} {58}},\
  \bibinfo {pages} {1490} (\bibinfo {year} {1987})}\BibitemShut {NoStop}%
\bibitem [{\citenamefont {Arnett}\ and\ \citenamefont
  {Rosner}(1987)}]{arnett1987neutrino}%
  \BibitemOpen
  \bibfield  {author} {\bibinfo {author} {\bibfnamefont {W.~D.}\ \bibnamefont
  {Arnett}}\ and\ \bibinfo {author} {\bibfnamefont {J.~L.}\ \bibnamefont
  {Rosner}},\ }\bibfield  {title} {\bibinfo {title} {Neutrino mass limits from
  sn1987a},\ }\href@noop {} {\bibfield  {journal} {\bibinfo  {journal}
  {Physical review letters}\ }\textbf {\bibinfo {volume} {58}},\ \bibinfo
  {pages} {1906} (\bibinfo {year} {1987})}\BibitemShut {NoStop}%
\bibitem [{\citenamefont {Lunardini}\ and\ \citenamefont
  {Smirnov}(2004)}]{lunardini2004neutrinos}%
  \BibitemOpen
  \bibfield  {author} {\bibinfo {author} {\bibfnamefont {C.}~\bibnamefont
  {Lunardini}}\ and\ \bibinfo {author} {\bibfnamefont {A.~Y.}\ \bibnamefont
  {Smirnov}},\ }\bibfield  {title} {\bibinfo {title} {Neutrinos from sn1987a:
  flavor conversion and interpretation of results},\ }\href@noop {} {\bibfield
  {journal} {\bibinfo  {journal} {Astroparticle Physics}\ }\textbf {\bibinfo
  {volume} {21}},\ \bibinfo {pages} {703} (\bibinfo {year} {2004})}\BibitemShut
  {NoStop}%
\bibitem [{\citenamefont {Blinnikov}\ \emph {et~al.}(1984)\citenamefont
  {Blinnikov}, \citenamefont {Novikov}, \citenamefont {Perevodchikova},\ and\
  \citenamefont {Polnarev}}]{blinnikov1984exploding}%
  \BibitemOpen
  \bibfield  {author} {\bibinfo {author} {\bibfnamefont {S.}~\bibnamefont
  {Blinnikov}}, \bibinfo {author} {\bibfnamefont {I.}~\bibnamefont {Novikov}},
  \bibinfo {author} {\bibfnamefont {T.}~\bibnamefont {Perevodchikova}},\ and\
  \bibinfo {author} {\bibfnamefont {A.}~\bibnamefont {Polnarev}},\ }\bibfield
  {title} {\bibinfo {title} {Exploding neutron stars in close binaries},\
  }\href@noop {} {\bibfield  {journal} {\bibinfo  {journal} {Soviet Astronomy
  Letters}\ }\textbf {\bibinfo {volume} {10}},\ \bibinfo {pages} {177}
  (\bibinfo {year} {1984})}\BibitemShut {NoStop}%
\bibitem [{\citenamefont {Paczynski}(1986)}]{paczynski1986gamma}%
  \BibitemOpen
  \bibfield  {author} {\bibinfo {author} {\bibfnamefont {B.}~\bibnamefont
  {Paczynski}},\ }\bibfield  {title} {\bibinfo {title} {Gamma-ray bursters at
  cosmological distances},\ }\href@noop {} {\bibfield  {journal} {\bibinfo
  {journal} {The Astrophysical Journal}\ }\textbf {\bibinfo {volume} {308}},\
  \bibinfo {pages} {L43} (\bibinfo {year} {1986})}\BibitemShut {NoStop}%
\bibitem [{\citenamefont {Eichler}\ \emph {et~al.}(1989)\citenamefont
  {Eichler}, \citenamefont {Livio}, \citenamefont {Piran},\ and\ \citenamefont
  {Schramm}}]{eichler1989nucleosynthesis}%
  \BibitemOpen
  \bibfield  {author} {\bibinfo {author} {\bibfnamefont {D.}~\bibnamefont
  {Eichler}}, \bibinfo {author} {\bibfnamefont {M.}~\bibnamefont {Livio}},
  \bibinfo {author} {\bibfnamefont {T.}~\bibnamefont {Piran}},\ and\ \bibinfo
  {author} {\bibfnamefont {D.~N.}\ \bibnamefont {Schramm}},\ }\bibfield
  {title} {\bibinfo {title} {Nucleosynthesis, neutrino bursts and $\gamma$-rays
  from coalescing neutron stars},\ }\href@noop {} {\bibfield  {journal}
  {\bibinfo  {journal} {Nature}\ }\textbf {\bibinfo {volume} {340}},\ \bibinfo
  {pages} {126} (\bibinfo {year} {1989})}\BibitemShut {NoStop}%
\bibitem [{\citenamefont {Abbott}\ \emph
  {et~al.}(2017{\natexlab{a}})\citenamefont {Abbott}, \citenamefont {Abbott},
  \citenamefont {Abbott}, \citenamefont {Acernese}, \citenamefont {Ackley},
  \citenamefont {Adams}, \citenamefont {Adams}, \citenamefont {Addesso},
  \citenamefont {Adhikari}, \citenamefont {Adya} \emph
  {et~al.}}]{abbott2017gw170817}%
  \BibitemOpen
  \bibfield  {author} {\bibinfo {author} {\bibfnamefont {B.~P.}\ \bibnamefont
  {Abbott}}, \bibinfo {author} {\bibfnamefont {R.}~\bibnamefont {Abbott}},
  \bibinfo {author} {\bibfnamefont {T.}~\bibnamefont {Abbott}}, \bibinfo
  {author} {\bibfnamefont {F.}~\bibnamefont {Acernese}}, \bibinfo {author}
  {\bibfnamefont {K.}~\bibnamefont {Ackley}}, \bibinfo {author} {\bibfnamefont
  {C.}~\bibnamefont {Adams}}, \bibinfo {author} {\bibfnamefont
  {T.}~\bibnamefont {Adams}}, \bibinfo {author} {\bibfnamefont
  {P.}~\bibnamefont {Addesso}}, \bibinfo {author} {\bibfnamefont
  {R.}~\bibnamefont {Adhikari}}, \bibinfo {author} {\bibfnamefont
  {V.}~\bibnamefont {Adya}}, \emph {et~al.},\ }\bibfield  {title} {\bibinfo
  {title} {Gw170817: observation of gravitational waves from a binary neutron
  star inspiral},\ }\href@noop {} {\bibfield  {journal} {\bibinfo  {journal}
  {Physical Review Letters}\ }\textbf {\bibinfo {volume} {119}},\ \bibinfo
  {pages} {161101} (\bibinfo {year} {2017}{\natexlab{a}})}\BibitemShut
  {NoStop}%
\bibitem [{\citenamefont {Abbott}\ \emph
  {et~al.}(2017{\natexlab{b}})\citenamefont {Abbott}, \citenamefont {Abbott},
  \citenamefont {Abbott}, \citenamefont {Acernese}, \citenamefont {Ackley},
  \citenamefont {Adams}, \citenamefont {Adams}, \citenamefont {Addesso},
  \citenamefont {Adhikari}, \citenamefont {Adya} \emph
  {et~al.}}]{abbott2017gravitational}%
  \BibitemOpen
  \bibfield  {author} {\bibinfo {author} {\bibfnamefont {B.~P.}\ \bibnamefont
  {Abbott}}, \bibinfo {author} {\bibfnamefont {R.}~\bibnamefont {Abbott}},
  \bibinfo {author} {\bibfnamefont {T.}~\bibnamefont {Abbott}}, \bibinfo
  {author} {\bibfnamefont {F.}~\bibnamefont {Acernese}}, \bibinfo {author}
  {\bibfnamefont {K.}~\bibnamefont {Ackley}}, \bibinfo {author} {\bibfnamefont
  {C.}~\bibnamefont {Adams}}, \bibinfo {author} {\bibfnamefont
  {T.}~\bibnamefont {Adams}}, \bibinfo {author} {\bibfnamefont
  {P.}~\bibnamefont {Addesso}}, \bibinfo {author} {\bibfnamefont
  {R.}~\bibnamefont {Adhikari}}, \bibinfo {author} {\bibfnamefont
  {V.}~\bibnamefont {Adya}}, \emph {et~al.},\ }\bibfield  {title} {\bibinfo
  {title} {Gravitational waves and gamma-rays from a binary neutron star
  merger: Gw170817 and grb 170817a},\ }\href@noop {} {\bibfield  {journal}
  {\bibinfo  {journal} {The Astrophysical Journal Letters}\ }\textbf {\bibinfo
  {volume} {848}},\ \bibinfo {pages} {L13} (\bibinfo {year}
  {2017}{\natexlab{b}})}\BibitemShut {NoStop}%
\bibitem [{\citenamefont {Goldstein}\ \emph {et~al.}(2017)\citenamefont
  {Goldstein}, \citenamefont {Veres}, \citenamefont {Burns}, \citenamefont
  {Briggs}, \citenamefont {Hamburg}, \citenamefont {Kocevski}, \citenamefont
  {Wilson-Hodge}, \citenamefont {Preece}, \citenamefont {Poolakkil},
  \citenamefont {Roberts} \emph {et~al.}}]{goldstein2017ordinary}%
  \BibitemOpen
  \bibfield  {author} {\bibinfo {author} {\bibfnamefont {A.}~\bibnamefont
  {Goldstein}}, \bibinfo {author} {\bibfnamefont {P.}~\bibnamefont {Veres}},
  \bibinfo {author} {\bibfnamefont {E.}~\bibnamefont {Burns}}, \bibinfo
  {author} {\bibfnamefont {M.}~\bibnamefont {Briggs}}, \bibinfo {author}
  {\bibfnamefont {R.}~\bibnamefont {Hamburg}}, \bibinfo {author} {\bibfnamefont
  {D.}~\bibnamefont {Kocevski}}, \bibinfo {author} {\bibfnamefont
  {C.}~\bibnamefont {Wilson-Hodge}}, \bibinfo {author} {\bibfnamefont
  {R.}~\bibnamefont {Preece}}, \bibinfo {author} {\bibfnamefont
  {S.}~\bibnamefont {Poolakkil}}, \bibinfo {author} {\bibfnamefont
  {O.}~\bibnamefont {Roberts}}, \emph {et~al.},\ }\bibfield  {title} {\bibinfo
  {title} {An ordinary short gamma-ray burst with extraordinary implications:
  Fermi-gbm detection of grb 170817a},\ }\href@noop {} {\bibfield  {journal}
  {\bibinfo  {journal} {The Astrophysical Journal Letters}\ }\textbf {\bibinfo
  {volume} {848}},\ \bibinfo {pages} {L14} (\bibinfo {year}
  {2017})}\BibitemShut {NoStop}%
\bibitem [{\citenamefont {Aasi}\ \emph {et~al.}(2014)\citenamefont {Aasi},
  \citenamefont {Abbott}, \citenamefont {Abbott}, \citenamefont {Abbott},
  \citenamefont {Abernathy}, \citenamefont {Acernese}, \citenamefont {Ackley},
  \citenamefont {Adams}, \citenamefont {Adams}, \citenamefont {Addesso} \emph
  {et~al.}}]{aasi2014methods}%
  \BibitemOpen
  \bibfield  {author} {\bibinfo {author} {\bibfnamefont {J.}~\bibnamefont
  {Aasi}}, \bibinfo {author} {\bibfnamefont {B.}~\bibnamefont {Abbott}},
  \bibinfo {author} {\bibfnamefont {R.}~\bibnamefont {Abbott}}, \bibinfo
  {author} {\bibfnamefont {T.}~\bibnamefont {Abbott}}, \bibinfo {author}
  {\bibfnamefont {M.~R.}\ \bibnamefont {Abernathy}}, \bibinfo {author}
  {\bibfnamefont {F.}~\bibnamefont {Acernese}}, \bibinfo {author}
  {\bibfnamefont {K.}~\bibnamefont {Ackley}}, \bibinfo {author} {\bibfnamefont
  {C.}~\bibnamefont {Adams}}, \bibinfo {author} {\bibfnamefont
  {T.}~\bibnamefont {Adams}}, \bibinfo {author} {\bibfnamefont
  {P.}~\bibnamefont {Addesso}}, \emph {et~al.},\ }\bibfield  {title} {\bibinfo
  {title} {Methods and results of a search for gravitational waves associated
  with gamma-ray bursts using the geo 600, ligo, and virgo detectors},\
  }\href@noop {} {\bibfield  {journal} {\bibinfo  {journal} {Physical Review
  D}\ }\textbf {\bibinfo {volume} {89}},\ \bibinfo {pages} {122004} (\bibinfo
  {year} {2014})}\BibitemShut {NoStop}%
\bibitem [{\citenamefont {Singer}\ \emph {et~al.}(2014)\citenamefont {Singer},
  \citenamefont {Price}, \citenamefont {Farr}, \citenamefont {Urban},
  \citenamefont {Pankow}, \citenamefont {Vitale}, \citenamefont {Veitch},
  \citenamefont {Farr}, \citenamefont {Hanna}, \citenamefont {Cannon} \emph
  {et~al.}}]{singer2014first}%
  \BibitemOpen
  \bibfield  {author} {\bibinfo {author} {\bibfnamefont {L.~P.}\ \bibnamefont
  {Singer}}, \bibinfo {author} {\bibfnamefont {L.~R.}\ \bibnamefont {Price}},
  \bibinfo {author} {\bibfnamefont {B.}~\bibnamefont {Farr}}, \bibinfo {author}
  {\bibfnamefont {A.~L.}\ \bibnamefont {Urban}}, \bibinfo {author}
  {\bibfnamefont {C.}~\bibnamefont {Pankow}}, \bibinfo {author} {\bibfnamefont
  {S.}~\bibnamefont {Vitale}}, \bibinfo {author} {\bibfnamefont
  {J.}~\bibnamefont {Veitch}}, \bibinfo {author} {\bibfnamefont {W.~M.}\
  \bibnamefont {Farr}}, \bibinfo {author} {\bibfnamefont {C.}~\bibnamefont
  {Hanna}}, \bibinfo {author} {\bibfnamefont {K.}~\bibnamefont {Cannon}}, \emph
  {et~al.},\ }\bibfield  {title} {\bibinfo {title} {The first two years of
  electromagnetic follow-up with advanced ligo and virgo},\ }\href@noop {}
  {\bibfield  {journal} {\bibinfo  {journal} {The Astrophysical Journal}\
  }\textbf {\bibinfo {volume} {795}},\ \bibinfo {pages} {105} (\bibinfo {year}
  {2014})}\BibitemShut {NoStop}%
\bibitem [{\citenamefont {Abbott}\ \emph
  {et~al.}(2017{\natexlab{c}})\citenamefont {Abbott}, \citenamefont {Abbott},
  \citenamefont {Abbott}, \citenamefont {Abernathy}, \citenamefont {Acernese},
  \citenamefont {Ackley}, \citenamefont {Adams}, \citenamefont {Adams},
  \citenamefont {Addesso}, \citenamefont {Adhikari} \emph
  {et~al.}}]{abbott2017search}%
  \BibitemOpen
  \bibfield  {author} {\bibinfo {author} {\bibfnamefont {B.~P.}\ \bibnamefont
  {Abbott}}, \bibinfo {author} {\bibfnamefont {R.}~\bibnamefont {Abbott}},
  \bibinfo {author} {\bibfnamefont {T.}~\bibnamefont {Abbott}}, \bibinfo
  {author} {\bibfnamefont {M.}~\bibnamefont {Abernathy}}, \bibinfo {author}
  {\bibfnamefont {F.}~\bibnamefont {Acernese}}, \bibinfo {author}
  {\bibfnamefont {K.}~\bibnamefont {Ackley}}, \bibinfo {author} {\bibfnamefont
  {C.}~\bibnamefont {Adams}}, \bibinfo {author} {\bibfnamefont
  {T.}~\bibnamefont {Adams}}, \bibinfo {author} {\bibfnamefont
  {P.}~\bibnamefont {Addesso}}, \bibinfo {author} {\bibfnamefont
  {R.}~\bibnamefont {Adhikari}}, \emph {et~al.},\ }\bibfield  {title} {\bibinfo
  {title} {Search for gravitational waves associated with gamma-ray bursts
  during the first advanced ligo observing run and implications for the origin
  of grb 150906b},\ }\href@noop {} {\bibfield  {journal} {\bibinfo  {journal}
  {The Astrophysical Journal}\ }\textbf {\bibinfo {volume} {841}},\ \bibinfo
  {pages} {89} (\bibinfo {year} {2017}{\natexlab{c}})}\BibitemShut {NoStop}%
\bibitem [{\citenamefont {Urban}(2016)}]{urban2016}%
  \BibitemOpen
  \bibfield  {author} {\bibinfo {author} {\bibfnamefont {A.~L.}\ \bibnamefont
  {Urban}},\ }\emph {\bibinfo {title} {Monsters in the Dark: High Energy
  Signatures of Black Hole Formation with Multimessenger Astronomy}},\
  \href@noop {} {Ph.D. thesis},\ \bibinfo  {school} {University of
  Wisconsin-Milwaukee} (\bibinfo {year} {2016})\BibitemShut {NoStop}%
\bibitem [{\citenamefont {Cho}(2019)}]{cho2019low}%
  \BibitemOpen
  \bibfield  {author} {\bibinfo {author} {\bibfnamefont {M.-A.}\ \bibnamefont
  {Cho}},\ }\emph {\bibinfo {title} {Low-Latency Searches for Gravitational
  Waves and Their Electromagnetic Counterparts with Advanced LIGO and Virgo}},\
  \href@noop {} {Ph.D. thesis},\ \bibinfo  {school} {University of Maryland}
  (\bibinfo {year} {2019})\BibitemShut {NoStop}%
\bibitem [{\citenamefont {Hamburg}\ \emph {et~al.}(2020)\citenamefont
  {Hamburg}, \citenamefont {Fletcher}, \citenamefont {Burns}, \citenamefont
  {Goldstein}, \citenamefont {Bissaldi}, \citenamefont {Briggs}, \citenamefont
  {Cleveland}, \citenamefont {Giles}, \citenamefont {Hui}, \citenamefont
  {Kocevski} \emph {et~al.}}]{hamburg2020joint}%
  \BibitemOpen
  \bibfield  {author} {\bibinfo {author} {\bibfnamefont {R.}~\bibnamefont
  {Hamburg}}, \bibinfo {author} {\bibfnamefont {C.}~\bibnamefont {Fletcher}},
  \bibinfo {author} {\bibfnamefont {E.}~\bibnamefont {Burns}}, \bibinfo
  {author} {\bibfnamefont {A.}~\bibnamefont {Goldstein}}, \bibinfo {author}
  {\bibfnamefont {E.}~\bibnamefont {Bissaldi}}, \bibinfo {author}
  {\bibfnamefont {M.}~\bibnamefont {Briggs}}, \bibinfo {author} {\bibfnamefont
  {W.}~\bibnamefont {Cleveland}}, \bibinfo {author} {\bibfnamefont
  {M.}~\bibnamefont {Giles}}, \bibinfo {author} {\bibfnamefont
  {C.}~\bibnamefont {Hui}}, \bibinfo {author} {\bibfnamefont {D.}~\bibnamefont
  {Kocevski}}, \emph {et~al.},\ }\bibfield  {title} {\bibinfo {title} {A joint
  fermi-gbm and ligo/virgo analysis of compact binary mergers from the first
  and second gravitational-wave observing runs},\ }\href@noop {} {\bibfield
  {journal} {\bibinfo  {journal} {The Astrophysical Journal}\ }\textbf
  {\bibinfo {volume} {893}},\ \bibinfo {pages} {100} (\bibinfo {year}
  {2020})}\BibitemShut {NoStop}%
\bibitem [{\citenamefont {Abbott}\ \emph {et~al.}(2020)\citenamefont {Abbott},
  \citenamefont {Abbott}, \citenamefont {Abraham}, \citenamefont {Acernese},
  \citenamefont {Ackley}, \citenamefont {Adams}, \citenamefont {Adhikari},
  \citenamefont {Adya}, \citenamefont {Affeldt}, \citenamefont {Agathos} \emph
  {et~al.}}]{abbott2020search}%
  \BibitemOpen
  \bibfield  {author} {\bibinfo {author} {\bibfnamefont {R.}~\bibnamefont
  {Abbott}}, \bibinfo {author} {\bibfnamefont {T.}~\bibnamefont {Abbott}},
  \bibinfo {author} {\bibfnamefont {S.}~\bibnamefont {Abraham}}, \bibinfo
  {author} {\bibfnamefont {F.}~\bibnamefont {Acernese}}, \bibinfo {author}
  {\bibfnamefont {K.}~\bibnamefont {Ackley}}, \bibinfo {author} {\bibfnamefont
  {C.}~\bibnamefont {Adams}}, \bibinfo {author} {\bibfnamefont
  {R.}~\bibnamefont {Adhikari}}, \bibinfo {author} {\bibfnamefont
  {V.}~\bibnamefont {Adya}}, \bibinfo {author} {\bibfnamefont {C.}~\bibnamefont
  {Affeldt}}, \bibinfo {author} {\bibfnamefont {M.}~\bibnamefont {Agathos}},
  \emph {et~al.},\ }\bibfield  {title} {\bibinfo {title} {Search for
  gravitational waves associated with gamma-ray bursts detected by fermi and
  swift during the ligo-virgo run o3a},\ }\href@noop {} {\bibfield  {journal}
  {\bibinfo  {journal} {arXiv preprint arXiv:2010.14550}\ } (\bibinfo {year}
  {2020})}\BibitemShut {NoStop}%
\bibitem [{\citenamefont {Stachie}\ \emph {et~al.}(2020)\citenamefont
  {Stachie}, \citenamefont {Dal~Canton}, \citenamefont {Burns}, \citenamefont
  {Christensen}, \citenamefont {Hamburg}, \citenamefont {Briggs}, \citenamefont
  {Broida}, \citenamefont {Goldstein}, \citenamefont {Hayes}, \citenamefont
  {Littenberg} \emph {et~al.}}]{Cosmin}%
  \BibitemOpen
  \bibfield  {author} {\bibinfo {author} {\bibfnamefont {C.}~\bibnamefont
  {Stachie}}, \bibinfo {author} {\bibfnamefont {T.}~\bibnamefont {Dal~Canton}},
  \bibinfo {author} {\bibfnamefont {E.}~\bibnamefont {Burns}}, \bibinfo
  {author} {\bibfnamefont {N.}~\bibnamefont {Christensen}}, \bibinfo {author}
  {\bibfnamefont {R.}~\bibnamefont {Hamburg}}, \bibinfo {author} {\bibfnamefont
  {M.}~\bibnamefont {Briggs}}, \bibinfo {author} {\bibfnamefont
  {J.}~\bibnamefont {Broida}}, \bibinfo {author} {\bibfnamefont
  {A.}~\bibnamefont {Goldstein}}, \bibinfo {author} {\bibfnamefont
  {F.}~\bibnamefont {Hayes}}, \bibinfo {author} {\bibfnamefont
  {T.}~\bibnamefont {Littenberg}}, \emph {et~al.},\ }\bibfield  {title}
  {\bibinfo {title} {Search for advanced ligo single interferometer compact
  binary coalescence signals in coincidence with gamma-ray events in
  fermi-gbm},\ }\href@noop {} {\bibfield  {journal} {\bibinfo  {journal}
  {Classical and Quantum Gravity}\ }\textbf {\bibinfo {volume} {37}},\ \bibinfo
  {pages} {175001} (\bibinfo {year} {2020})}\BibitemShut {NoStop}%
\bibitem [{\citenamefont {Nitz}\ \emph {et~al.}(2019)\citenamefont {Nitz},
  \citenamefont {Nielsen},\ and\ \citenamefont {Capano}}]{nitz2019potential}%
  \BibitemOpen
  \bibfield  {author} {\bibinfo {author} {\bibfnamefont {A.~H.}\ \bibnamefont
  {Nitz}}, \bibinfo {author} {\bibfnamefont {A.~B.}\ \bibnamefont {Nielsen}},\
  and\ \bibinfo {author} {\bibfnamefont {C.~D.}\ \bibnamefont {Capano}},\
  }\bibfield  {title} {\bibinfo {title} {Potential gravitational-wave and
  gamma-ray multi-messenger candidate from 2015 october 30},\ }\href@noop {}
  {\bibfield  {journal} {\bibinfo  {journal} {The Astrophysical Journal
  Letters}\ }\textbf {\bibinfo {volume} {876}},\ \bibinfo {pages} {L4}
  (\bibinfo {year} {2019})}\BibitemShut {NoStop}%
\bibitem [{\citenamefont {Abbott}\ \emph {et~al.}(2018)\citenamefont {Abbott},
  \citenamefont {Abbott}, \citenamefont {Abbott}, \citenamefont {Acernese},
  \citenamefont {Ackley}, \citenamefont {Adams}, \citenamefont {Adams},
  \citenamefont {Addesso}, \citenamefont {Adhikari}, \citenamefont {Adya} \emph
  {et~al.}}]{abbott2018gw170817}%
  \BibitemOpen
  \bibfield  {author} {\bibinfo {author} {\bibfnamefont {B.}~\bibnamefont
  {Abbott}}, \bibinfo {author} {\bibfnamefont {R.}~\bibnamefont {Abbott}},
  \bibinfo {author} {\bibfnamefont {T.}~\bibnamefont {Abbott}}, \bibinfo
  {author} {\bibfnamefont {F.}~\bibnamefont {Acernese}}, \bibinfo {author}
  {\bibfnamefont {K.}~\bibnamefont {Ackley}}, \bibinfo {author} {\bibfnamefont
  {C.}~\bibnamefont {Adams}}, \bibinfo {author} {\bibfnamefont
  {T.}~\bibnamefont {Adams}}, \bibinfo {author} {\bibfnamefont
  {P.}~\bibnamefont {Addesso}}, \bibinfo {author} {\bibfnamefont
  {R.}~\bibnamefont {Adhikari}}, \bibinfo {author} {\bibfnamefont
  {V.}~\bibnamefont {Adya}}, \emph {et~al.},\ }\bibfield  {title} {\bibinfo
  {title} {Gw170817: Measurements of neutron star radii and equation of
  state},\ }\href@noop {} {\bibfield  {journal} {\bibinfo  {journal} {Physical
  review letters}\ }\textbf {\bibinfo {volume} {121}},\ \bibinfo {pages}
  {161101} (\bibinfo {year} {2018})}\BibitemShut {NoStop}%
\bibitem [{\citenamefont {Coughlin}\ \emph {et~al.}(2019)\citenamefont
  {Coughlin}, \citenamefont {Dietrich}, \citenamefont {Margalit},\ and\
  \citenamefont {Metzger}}]{coughlin2019multimessenger}%
  \BibitemOpen
  \bibfield  {author} {\bibinfo {author} {\bibfnamefont {M.~W.}\ \bibnamefont
  {Coughlin}}, \bibinfo {author} {\bibfnamefont {T.}~\bibnamefont {Dietrich}},
  \bibinfo {author} {\bibfnamefont {B.}~\bibnamefont {Margalit}},\ and\
  \bibinfo {author} {\bibfnamefont {B.~D.}\ \bibnamefont {Metzger}},\
  }\bibfield  {title} {\bibinfo {title} {Multimessenger bayesian parameter
  inference of a binary neutron star merger},\ }\href@noop {} {\bibfield
  {journal} {\bibinfo  {journal} {Monthly Notices of the Royal Astronomical
  Society: Letters}\ }\textbf {\bibinfo {volume} {489}},\ \bibinfo {pages}
  {L91} (\bibinfo {year} {2019})}\BibitemShut {NoStop}%
\bibitem [{\citenamefont {Collaboration}\ \emph {et~al.}(2017)\citenamefont
  {Collaboration}, \citenamefont {Collaboration}, \citenamefont
  {Collaboration}, \citenamefont {Collaboration}, \citenamefont
  {Collaboration}, \citenamefont {Collaboration}, \citenamefont
  {Collaboration}, \citenamefont {Collaboration}, \citenamefont {Collaboration}
  \emph {et~al.}}]{ligo2017gravitational}%
  \BibitemOpen
  \bibfield  {author} {\bibinfo {author} {\bibfnamefont {L.~S.}\ \bibnamefont
  {Collaboration}}, \bibinfo {author} {\bibfnamefont {V.}~\bibnamefont
  {Collaboration}}, \bibinfo {author} {\bibfnamefont {M.}~\bibnamefont
  {Collaboration}}, \bibinfo {author} {\bibfnamefont {D.~E. C. G.-E.}\
  \bibnamefont {Collaboration}}, \bibinfo {author} {\bibfnamefont
  {D.}~\bibnamefont {Collaboration}}, \bibinfo {author} {\bibfnamefont
  {D.}~\bibnamefont {Collaboration}}, \bibinfo {author} {\bibfnamefont
  {L.~C.~O.}\ \bibnamefont {Collaboration}}, \bibinfo {author} {\bibfnamefont
  {V.}~\bibnamefont {Collaboration}}, \bibinfo {author} {\bibfnamefont
  {M.}~\bibnamefont {Collaboration}}, \emph {et~al.},\ }\bibfield  {title}
  {\bibinfo {title} {A gravitational-wave standard siren measurement of the
  hubble constant},\ }\href@noop {} {\bibfield  {journal} {\bibinfo  {journal}
  {Nature}\ }\textbf {\bibinfo {volume} {551}},\ \bibinfo {pages} {85}
  (\bibinfo {year} {2017})}\BibitemShut {NoStop}%
\bibitem [{\citenamefont {Farah}\ \emph {et~al.}(2020)\citenamefont {Farah},
  \citenamefont {Essick}, \citenamefont {Doctor}, \citenamefont {Fishbach},\
  and\ \citenamefont {Holz}}]{farah2020counting}%
  \BibitemOpen
  \bibfield  {author} {\bibinfo {author} {\bibfnamefont {A.}~\bibnamefont
  {Farah}}, \bibinfo {author} {\bibfnamefont {R.}~\bibnamefont {Essick}},
  \bibinfo {author} {\bibfnamefont {Z.}~\bibnamefont {Doctor}}, \bibinfo
  {author} {\bibfnamefont {M.}~\bibnamefont {Fishbach}},\ and\ \bibinfo
  {author} {\bibfnamefont {D.~E.}\ \bibnamefont {Holz}},\ }\bibfield  {title}
  {\bibinfo {title} {Counting on short gamma-ray bursts: Gravitational-wave
  constraints of jet geometry},\ }\href@noop {} {\bibfield  {journal} {\bibinfo
   {journal} {The Astrophysical Journal}\ }\textbf {\bibinfo {volume} {895}},\
  \bibinfo {pages} {108} (\bibinfo {year} {2020})}\BibitemShut {NoStop}%
\bibitem [{\citenamefont {Ashton}\ \emph {et~al.}(2018)\citenamefont {Ashton},
  \citenamefont {Burns}, \citenamefont {Dal~Canton}, \citenamefont {Dent},
  \citenamefont {Eggenstein}, \citenamefont {Nielsen}, \citenamefont {Prix},
  \citenamefont {Was},\ and\ \citenamefont {Zhu}}]{Ashton}%
  \BibitemOpen
  \bibfield  {author} {\bibinfo {author} {\bibfnamefont {G.}~\bibnamefont
  {Ashton}}, \bibinfo {author} {\bibfnamefont {E.}~\bibnamefont {Burns}},
  \bibinfo {author} {\bibfnamefont {T.}~\bibnamefont {Dal~Canton}}, \bibinfo
  {author} {\bibfnamefont {T.}~\bibnamefont {Dent}}, \bibinfo {author}
  {\bibfnamefont {H.-B.}\ \bibnamefont {Eggenstein}}, \bibinfo {author}
  {\bibfnamefont {A.~B.}\ \bibnamefont {Nielsen}}, \bibinfo {author}
  {\bibfnamefont {R.}~\bibnamefont {Prix}}, \bibinfo {author} {\bibfnamefont
  {M.}~\bibnamefont {Was}},\ and\ \bibinfo {author} {\bibfnamefont {S.~J.}\
  \bibnamefont {Zhu}},\ }\bibfield  {title} {\bibinfo {title} {Coincident
  detection significance in multimessenger astronomy},\ }\href@noop {}
  {\bibfield  {journal} {\bibinfo  {journal} {The Astrophysical Journal}\
  }\textbf {\bibinfo {volume} {860}},\ \bibinfo {pages} {6} (\bibinfo {year}
  {2018})}\BibitemShut {NoStop}%
\bibitem [{\citenamefont {Burns}(2019)}]{burns2019neutron}%
  \BibitemOpen
  \bibfield  {author} {\bibinfo {author} {\bibfnamefont {E.}~\bibnamefont
  {Burns}},\ }\bibfield  {title} {\bibinfo {title} {Neutron star mergers and
  how to study them},\ }\href@noop {} {\bibfield  {journal} {\bibinfo
  {journal} {arXiv preprint arXiv:1909.06085}\ } (\bibinfo {year}
  {2019})}\BibitemShut {NoStop}%
\bibitem [{\citenamefont {Bartos}\ \emph {et~al.}(2019)\citenamefont {Bartos},
  \citenamefont {Veske}, \citenamefont {Keivani}, \citenamefont {M{\'a}rka},
  \citenamefont {Countryman}, \citenamefont {Blaufuss}, \citenamefont
  {Finley},\ and\ \citenamefont {M{\'a}rka}}]{bartos2019bayesian}%
  \BibitemOpen
  \bibfield  {author} {\bibinfo {author} {\bibfnamefont {I.}~\bibnamefont
  {Bartos}}, \bibinfo {author} {\bibfnamefont {D.}~\bibnamefont {Veske}},
  \bibinfo {author} {\bibfnamefont {A.}~\bibnamefont {Keivani}}, \bibinfo
  {author} {\bibfnamefont {Z.}~\bibnamefont {M{\'a}rka}}, \bibinfo {author}
  {\bibfnamefont {S.}~\bibnamefont {Countryman}}, \bibinfo {author}
  {\bibfnamefont {E.}~\bibnamefont {Blaufuss}}, \bibinfo {author}
  {\bibfnamefont {C.}~\bibnamefont {Finley}},\ and\ \bibinfo {author}
  {\bibfnamefont {S.}~\bibnamefont {M{\'a}rka}},\ }\bibfield  {title} {\bibinfo
  {title} {Bayesian multimessenger search method for common sources of
  gravitational waves and high-energy neutrinos},\ }\href@noop {} {\bibfield
  {journal} {\bibinfo  {journal} {Physical Review D}\ }\textbf {\bibinfo
  {volume} {100}},\ \bibinfo {pages} {083017} (\bibinfo {year}
  {2019})}\BibitemShut {NoStop}%
\bibitem [{\citenamefont {Veske}\ \emph {et~al.}(2020)\citenamefont {Veske},
  \citenamefont {M{\'a}rka}, \citenamefont {Bartos},\ and\ \citenamefont
  {M{\'a}rka}}]{veske2020search}%
  \BibitemOpen
  \bibfield  {author} {\bibinfo {author} {\bibfnamefont {D.}~\bibnamefont
  {Veske}}, \bibinfo {author} {\bibfnamefont {Z.}~\bibnamefont {M{\'a}rka}},
  \bibinfo {author} {\bibfnamefont {I.}~\bibnamefont {Bartos}},\ and\ \bibinfo
  {author} {\bibfnamefont {S.}~\bibnamefont {M{\'a}rka}},\ }\bibfield  {title}
  {\bibinfo {title} {How to search for multiple messengers--a general framework
  beyond two messengers},\ }\href@noop {} {\bibfield  {journal} {\bibinfo
  {journal} {arXiv preprint arXiv:2010.04162}\ } (\bibinfo {year}
  {2020})}\BibitemShut {NoStop}%
\bibitem [{\citenamefont {Ashton}\ \emph {et~al.}(2020)\citenamefont {Ashton},
  \citenamefont {Ackley}, \citenamefont {Hernandez},\ and\ \citenamefont
  {Piotrzkowski}}]{ashton2020current}%
  \BibitemOpen
  \bibfield  {author} {\bibinfo {author} {\bibfnamefont {G.}~\bibnamefont
  {Ashton}}, \bibinfo {author} {\bibfnamefont {K.}~\bibnamefont {Ackley}},
  \bibinfo {author} {\bibfnamefont {I.~M.}\ \bibnamefont {Hernandez}},\ and\
  \bibinfo {author} {\bibfnamefont {B.}~\bibnamefont {Piotrzkowski}},\
  }\bibfield  {title} {\bibinfo {title} {Current observations are insufficient
  to confidently associate the binary black hole merger gw190521 with agn
  j124942. 3+ 344929},\ }\href@noop {} {\bibfield  {journal} {\bibinfo
  {journal} {arXiv preprint arXiv:2009.12346}\ } (\bibinfo {year}
  {2020})}\BibitemShut {NoStop}%
\bibitem [{\citenamefont {Howell}\ \emph {et~al.}(2019)\citenamefont {Howell},
  \citenamefont {Ackley}, \citenamefont {Rowlinson},\ and\ \citenamefont
  {Coward}}]{howell2019joint}%
  \BibitemOpen
  \bibfield  {author} {\bibinfo {author} {\bibfnamefont {E.}~\bibnamefont
  {Howell}}, \bibinfo {author} {\bibfnamefont {K.}~\bibnamefont {Ackley}},
  \bibinfo {author} {\bibfnamefont {A.}~\bibnamefont {Rowlinson}},\ and\
  \bibinfo {author} {\bibfnamefont {D.}~\bibnamefont {Coward}},\ }\bibfield
  {title} {\bibinfo {title} {Joint gravitational wave--gamma-ray burst
  detection rates in the aftermath of gw170817},\ }\href@noop {} {\bibfield
  {journal} {\bibinfo  {journal} {Monthly Notices of the Royal Astronomical
  Society}\ }\textbf {\bibinfo {volume} {485}},\ \bibinfo {pages} {1435}
  (\bibinfo {year} {2019})}\BibitemShut {NoStop}%
\bibitem [{\citenamefont {{G{\'o}rski}}\ \emph {et~al.}(2005)\citenamefont
  {{G{\'o}rski}}, \citenamefont {{Hivon}}, \citenamefont {{Banday}},
  \citenamefont {{Wand elt}}, \citenamefont {{Hansen}}, \citenamefont
  {{Reinecke}},\ and\ \citenamefont {{Bartelmann}}}]{HEALPix}%
  \BibitemOpen
  \bibfield  {author} {\bibinfo {author} {\bibfnamefont {K.~M.}\ \bibnamefont
  {{G{\'o}rski}}}, \bibinfo {author} {\bibfnamefont {E.}~\bibnamefont
  {{Hivon}}}, \bibinfo {author} {\bibfnamefont {A.~J.}\ \bibnamefont
  {{Banday}}}, \bibinfo {author} {\bibfnamefont {B.~D.}\ \bibnamefont {{Wand
  elt}}}, \bibinfo {author} {\bibfnamefont {F.~K.}\ \bibnamefont {{Hansen}}},
  \bibinfo {author} {\bibfnamefont {M.}~\bibnamefont {{Reinecke}}},\ and\
  \bibinfo {author} {\bibfnamefont {M.}~\bibnamefont {{Bartelmann}}},\
  }\bibfield  {title} {\bibinfo {title} {{HEALPix: A Framework for
  High-Resolution Discretization and Fast Analysis of Data Distributed on the
  Sphere}},\ }\href {https://doi.org/10.1086/427976} {\bibfield  {journal}
  {\bibinfo  {journal} {\apj}\ }\textbf {\bibinfo {volume} {622}},\ \bibinfo
  {pages} {759} (\bibinfo {year} {2005})},\ \Eprint
  {https://arxiv.org/abs/astro-ph/0409513} {arXiv:astro-ph/0409513 [astro-ph]}
  \BibitemShut {NoStop}%
\bibitem [{\citenamefont {Singer}\ and\ \citenamefont
  {Price}(2016)}]{singer2016rapid}%
  \BibitemOpen
  \bibfield  {author} {\bibinfo {author} {\bibfnamefont {L.~P.}\ \bibnamefont
  {Singer}}\ and\ \bibinfo {author} {\bibfnamefont {L.~R.}\ \bibnamefont
  {Price}},\ }\bibfield  {title} {\bibinfo {title} {Rapid bayesian position
  reconstruction for gravitational-wave transients},\ }\href@noop {} {\bibfield
   {journal} {\bibinfo  {journal} {Physical Review D}\ }\textbf {\bibinfo
  {volume} {93}},\ \bibinfo {pages} {024013} (\bibinfo {year}
  {2016})}\BibitemShut {NoStop}%
\bibitem [{\citenamefont {D{\'a}lya}\ \emph {et~al.}(2018)\citenamefont
  {D{\'a}lya}, \citenamefont {Galg{\'o}czi}, \citenamefont {Dobos},
  \citenamefont {Frei}, \citenamefont {Heng}, \citenamefont {Macas},
  \citenamefont {Messenger}, \citenamefont {Raffai},\ and\ \citenamefont
  {de~Souza}}]{dalya2018glade}%
  \BibitemOpen
  \bibfield  {author} {\bibinfo {author} {\bibfnamefont {G.}~\bibnamefont
  {D{\'a}lya}}, \bibinfo {author} {\bibfnamefont {G.}~\bibnamefont
  {Galg{\'o}czi}}, \bibinfo {author} {\bibfnamefont {L.}~\bibnamefont {Dobos}},
  \bibinfo {author} {\bibfnamefont {Z.}~\bibnamefont {Frei}}, \bibinfo {author}
  {\bibfnamefont {I.~S.}\ \bibnamefont {Heng}}, \bibinfo {author}
  {\bibfnamefont {R.}~\bibnamefont {Macas}}, \bibinfo {author} {\bibfnamefont
  {C.}~\bibnamefont {Messenger}}, \bibinfo {author} {\bibfnamefont
  {P.}~\bibnamefont {Raffai}},\ and\ \bibinfo {author} {\bibfnamefont {R.~S.}\
  \bibnamefont {de~Souza}},\ }\bibfield  {title} {\bibinfo {title} {Glade: A
  galaxy catalogue for multimessenger searches in the advanced
  gravitational-wave detector era},\ }\href@noop {} {\bibfield  {journal}
  {\bibinfo  {journal} {Monthly Notices of the Royal Astronomical Society}\
  }\textbf {\bibinfo {volume} {479}},\ \bibinfo {pages} {2374} (\bibinfo {year}
  {2018})}\BibitemShut {NoStop}%
\bibitem [{\citenamefont {Singer}\ \emph {et~al.}(2016)\citenamefont {Singer},
  \citenamefont {Chen}, \citenamefont {Holz}, \citenamefont {Farr},
  \citenamefont {Price}, \citenamefont {Raymond}, \citenamefont {Cenko},
  \citenamefont {Gehrels}, \citenamefont {Cannizzo}, \citenamefont {Kasliwal}
  \emph {et~al.}}]{singer2016supplement}%
  \BibitemOpen
  \bibfield  {author} {\bibinfo {author} {\bibfnamefont {L.~P.}\ \bibnamefont
  {Singer}}, \bibinfo {author} {\bibfnamefont {H.-Y.}\ \bibnamefont {Chen}},
  \bibinfo {author} {\bibfnamefont {D.~E.}\ \bibnamefont {Holz}}, \bibinfo
  {author} {\bibfnamefont {W.~M.}\ \bibnamefont {Farr}}, \bibinfo {author}
  {\bibfnamefont {L.~R.}\ \bibnamefont {Price}}, \bibinfo {author}
  {\bibfnamefont {V.}~\bibnamefont {Raymond}}, \bibinfo {author} {\bibfnamefont
  {S.~B.}\ \bibnamefont {Cenko}}, \bibinfo {author} {\bibfnamefont
  {N.}~\bibnamefont {Gehrels}}, \bibinfo {author} {\bibfnamefont
  {J.}~\bibnamefont {Cannizzo}}, \bibinfo {author} {\bibfnamefont {M.~M.}\
  \bibnamefont {Kasliwal}}, \emph {et~al.},\ }\bibfield  {title} {\bibinfo
  {title} {Supplement:“going the distance: Mapping host galaxies of ligo and
  virgo sources in three dimensions using local cosmography and targeted
  follow-up”(2016, apjl, 829, l15)},\ }\href@noop {} {\bibfield  {journal}
  {\bibinfo  {journal} {The Astrophysical Journal Supplement Series}\ }\textbf
  {\bibinfo {volume} {226}},\ \bibinfo {pages} {10} (\bibinfo {year}
  {2016})}\BibitemShut {NoStop}%
\bibitem [{\citenamefont {Fishbach}\ \emph {et~al.}(2019)\citenamefont
  {Fishbach}, \citenamefont {Gray}, \citenamefont {Hernandez}, \citenamefont
  {Qi}, \citenamefont {Sur}, \citenamefont {Acernese}, \citenamefont {Aiello},
  \citenamefont {Allocca}, \citenamefont {Aloy}, \citenamefont {Amato} \emph
  {et~al.}}]{fishbach2019standard}%
  \BibitemOpen
  \bibfield  {author} {\bibinfo {author} {\bibfnamefont {M.}~\bibnamefont
  {Fishbach}}, \bibinfo {author} {\bibfnamefont {R.}~\bibnamefont {Gray}},
  \bibinfo {author} {\bibfnamefont {I.~M.}\ \bibnamefont {Hernandez}}, \bibinfo
  {author} {\bibfnamefont {H.}~\bibnamefont {Qi}}, \bibinfo {author}
  {\bibfnamefont {A.}~\bibnamefont {Sur}}, \bibinfo {author} {\bibfnamefont
  {F.}~\bibnamefont {Acernese}}, \bibinfo {author} {\bibfnamefont
  {L.}~\bibnamefont {Aiello}}, \bibinfo {author} {\bibfnamefont
  {A.}~\bibnamefont {Allocca}}, \bibinfo {author} {\bibfnamefont
  {M.}~\bibnamefont {Aloy}}, \bibinfo {author} {\bibfnamefont {A.}~\bibnamefont
  {Amato}}, \emph {et~al.},\ }\bibfield  {title} {\bibinfo {title} {A standard
  siren measurement of the hubble constant from gw170817 without the
  electromagnetic counterpart},\ }\href@noop {} {\bibfield  {journal} {\bibinfo
   {journal} {The Astrophysical Journal Letters}\ }\textbf {\bibinfo {volume}
  {871}},\ \bibinfo {pages} {L13} (\bibinfo {year} {2019})}\BibitemShut
  {NoStop}%
\bibitem [{\citenamefont {Gehrels}\ \emph {et~al.}(2004)\citenamefont
  {Gehrels}, \citenamefont {Chincarini}, \citenamefont {Giommi}, \citenamefont
  {Mason}, \citenamefont {Nousek}, \citenamefont {Wells}, \citenamefont
  {White}, \citenamefont {Barthelmy}, \citenamefont {Burrows}, \citenamefont
  {Cominsky} \emph {et~al.}}]{gehrels2004swift}%
  \BibitemOpen
  \bibfield  {author} {\bibinfo {author} {\bibfnamefont {N.}~\bibnamefont
  {Gehrels}}, \bibinfo {author} {\bibfnamefont {G.}~\bibnamefont {Chincarini}},
  \bibinfo {author} {\bibfnamefont {P.}~\bibnamefont {Giommi}}, \bibinfo
  {author} {\bibfnamefont {K.}~\bibnamefont {Mason}}, \bibinfo {author}
  {\bibfnamefont {J.~A.}\ \bibnamefont {Nousek}}, \bibinfo {author}
  {\bibfnamefont {A.}~\bibnamefont {Wells}}, \bibinfo {author} {\bibfnamefont
  {N.}~\bibnamefont {White}}, \bibinfo {author} {\bibfnamefont
  {S.}~\bibnamefont {Barthelmy}}, \bibinfo {author} {\bibfnamefont {D.~N.}\
  \bibnamefont {Burrows}}, \bibinfo {author} {\bibfnamefont {L.}~\bibnamefont
  {Cominsky}}, \emph {et~al.},\ }\bibfield  {title} {\bibinfo {title} {The
  swift gamma-ray burst mission},\ }\href@noop {} {\bibfield  {journal}
  {\bibinfo  {journal} {The Astrophysical Journal}\ }\textbf {\bibinfo {volume}
  {611}},\ \bibinfo {pages} {1005} (\bibinfo {year} {2004})}\BibitemShut
  {NoStop}%
\bibitem [{\citenamefont {Valenti}\ \emph {et~al.}(2017)\citenamefont
  {Valenti}, \citenamefont {David}, \citenamefont {Yang}, \citenamefont
  {Cappellaro}, \citenamefont {Tartaglia}, \citenamefont {Corsi}, \citenamefont
  {Jha}, \citenamefont {Reichart}, \citenamefont {Haislip},\ and\ \citenamefont
  {Kouprianov}}]{valenti2017discovery}%
  \BibitemOpen
  \bibfield  {author} {\bibinfo {author} {\bibfnamefont {S.}~\bibnamefont
  {Valenti}}, \bibinfo {author} {\bibfnamefont {J.}~\bibnamefont {David}},
  \bibinfo {author} {\bibfnamefont {S.}~\bibnamefont {Yang}}, \bibinfo {author}
  {\bibfnamefont {E.}~\bibnamefont {Cappellaro}}, \bibinfo {author}
  {\bibfnamefont {L.}~\bibnamefont {Tartaglia}}, \bibinfo {author}
  {\bibfnamefont {A.}~\bibnamefont {Corsi}}, \bibinfo {author} {\bibfnamefont
  {S.~W.}\ \bibnamefont {Jha}}, \bibinfo {author} {\bibfnamefont {D.~E.}\
  \bibnamefont {Reichart}}, \bibinfo {author} {\bibfnamefont {J.}~\bibnamefont
  {Haislip}},\ and\ \bibinfo {author} {\bibfnamefont {V.}~\bibnamefont
  {Kouprianov}},\ }\bibfield  {title} {\bibinfo {title} {The discovery of the
  electromagnetic counterpart of gw170817: kilonova at 2017gfo/dlt17ck},\
  }\href@noop {} {\bibfield  {journal} {\bibinfo  {journal} {The Astrophysical
  Journal Letters}\ }\textbf {\bibinfo {volume} {848}},\ \bibinfo {pages} {L24}
  (\bibinfo {year} {2017})}\BibitemShut {NoStop}%
\bibitem [{\citenamefont {Abbott}\ \emph
  {et~al.}(2017{\natexlab{d}})\citenamefont {Abbott}, \citenamefont {Bloemen},
  \citenamefont {Canizares}, \citenamefont {Falcke}, \citenamefont {Fender},
  \citenamefont {Ghosh}, \citenamefont {Groot}, \citenamefont {Hinderer},
  \citenamefont {H{\"o}randel}, \citenamefont {Jonker} \emph
  {et~al.}}]{abbott2017multi}%
  \BibitemOpen
  \bibfield  {author} {\bibinfo {author} {\bibfnamefont {B.~P.}\ \bibnamefont
  {Abbott}}, \bibinfo {author} {\bibfnamefont {S.}~\bibnamefont {Bloemen}},
  \bibinfo {author} {\bibfnamefont {P.}~\bibnamefont {Canizares}}, \bibinfo
  {author} {\bibfnamefont {H.}~\bibnamefont {Falcke}}, \bibinfo {author}
  {\bibfnamefont {R.}~\bibnamefont {Fender}}, \bibinfo {author} {\bibfnamefont
  {S.}~\bibnamefont {Ghosh}}, \bibinfo {author} {\bibfnamefont
  {P.}~\bibnamefont {Groot}}, \bibinfo {author} {\bibfnamefont
  {T.}~\bibnamefont {Hinderer}}, \bibinfo {author} {\bibfnamefont
  {J.}~\bibnamefont {H{\"o}randel}}, \bibinfo {author} {\bibfnamefont
  {P.}~\bibnamefont {Jonker}}, \emph {et~al.},\ }\bibfield  {title} {\bibinfo
  {title} {Multi-messenger observations of a binary neutron star merger},\
  }\href@noop {} {\bibfield  {journal} {\bibinfo  {journal} {\apjl}\ }
  (\bibinfo {year} {2017}{\natexlab{d}})}\BibitemShut {NoStop}%
\bibitem [{\citenamefont {Abbott}\ \emph {et~al.}(2019)\citenamefont {Abbott},
  \citenamefont {Abbott}, \citenamefont {Abbott}, \citenamefont {Acernese},
  \citenamefont {Ackley}, \citenamefont {Adams}, \citenamefont {Adams},
  \citenamefont {Addesso}, \citenamefont {Adhikari}, \citenamefont {Adya} \emph
  {et~al.}}]{abbott2019properties}%
  \BibitemOpen
  \bibfield  {author} {\bibinfo {author} {\bibfnamefont {B.}~\bibnamefont
  {Abbott}}, \bibinfo {author} {\bibfnamefont {R.}~\bibnamefont {Abbott}},
  \bibinfo {author} {\bibfnamefont {T.}~\bibnamefont {Abbott}}, \bibinfo
  {author} {\bibfnamefont {F.}~\bibnamefont {Acernese}}, \bibinfo {author}
  {\bibfnamefont {K.}~\bibnamefont {Ackley}}, \bibinfo {author} {\bibfnamefont
  {C.}~\bibnamefont {Adams}}, \bibinfo {author} {\bibfnamefont
  {T.}~\bibnamefont {Adams}}, \bibinfo {author} {\bibfnamefont
  {P.}~\bibnamefont {Addesso}}, \bibinfo {author} {\bibfnamefont
  {R.}~\bibnamefont {Adhikari}}, \bibinfo {author} {\bibfnamefont
  {V.}~\bibnamefont {Adya}}, \emph {et~al.},\ }\bibfield  {title} {\bibinfo
  {title} {Properties of the binary neutron star merger gw170817},\ }\href@noop
  {} {\bibfield  {journal} {\bibinfo  {journal} {Physical Review X}\ }\textbf
  {\bibinfo {volume} {9}},\ \bibinfo {pages} {011001} (\bibinfo {year}
  {2019})}\BibitemShut {NoStop}%
\bibitem [{\citenamefont {von Kienlin}\ \emph {et~al.}(2020)\citenamefont {von
  Kienlin}, \citenamefont {Meegan}, \citenamefont {Paciesas}, \citenamefont
  {Bhat}, \citenamefont {Bissaldi}, \citenamefont {Briggs}, \citenamefont
  {Burns}, \citenamefont {Cleveland}, \citenamefont {Gibby}, \citenamefont
  {Giles} \emph {et~al.}}]{von2020fourth}%
  \BibitemOpen
  \bibfield  {author} {\bibinfo {author} {\bibfnamefont {A.}~\bibnamefont {von
  Kienlin}}, \bibinfo {author} {\bibfnamefont {C.}~\bibnamefont {Meegan}},
  \bibinfo {author} {\bibfnamefont {W.}~\bibnamefont {Paciesas}}, \bibinfo
  {author} {\bibfnamefont {P.}~\bibnamefont {Bhat}}, \bibinfo {author}
  {\bibfnamefont {E.}~\bibnamefont {Bissaldi}}, \bibinfo {author}
  {\bibfnamefont {M.}~\bibnamefont {Briggs}}, \bibinfo {author} {\bibfnamefont
  {E.}~\bibnamefont {Burns}}, \bibinfo {author} {\bibfnamefont
  {W.}~\bibnamefont {Cleveland}}, \bibinfo {author} {\bibfnamefont
  {M.}~\bibnamefont {Gibby}}, \bibinfo {author} {\bibfnamefont
  {M.}~\bibnamefont {Giles}}, \emph {et~al.},\ }\bibfield  {title} {\bibinfo
  {title} {The fourth fermi-gbm gamma-ray burst catalog: A decade of data},\
  }\href@noop {} {\bibfield  {journal} {\bibinfo  {journal} {The Astrophysical
  Journal}\ }\textbf {\bibinfo {volume} {893}},\ \bibinfo {pages} {46}
  (\bibinfo {year} {2020})}\BibitemShut {NoStop}%
\bibitem [{\citenamefont {Connaughton}\ \emph {et~al.}(2016)\citenamefont
  {Connaughton}, \citenamefont {Burns}, \citenamefont {Goldstein},
  \citenamefont {Blackburn}, \citenamefont {Briggs}, \citenamefont {Zhang},
  \citenamefont {Camp}, \citenamefont {Christensen}, \citenamefont {Hui},
  \citenamefont {Jenke} \emph {et~al.}}]{connaughton2016fermi}%
  \BibitemOpen
  \bibfield  {author} {\bibinfo {author} {\bibfnamefont {V.}~\bibnamefont
  {Connaughton}}, \bibinfo {author} {\bibfnamefont {E.}~\bibnamefont {Burns}},
  \bibinfo {author} {\bibfnamefont {A.}~\bibnamefont {Goldstein}}, \bibinfo
  {author} {\bibfnamefont {L.}~\bibnamefont {Blackburn}}, \bibinfo {author}
  {\bibfnamefont {M.}~\bibnamefont {Briggs}}, \bibinfo {author} {\bibfnamefont
  {B.-B.}\ \bibnamefont {Zhang}}, \bibinfo {author} {\bibfnamefont
  {J.}~\bibnamefont {Camp}}, \bibinfo {author} {\bibfnamefont {N.}~\bibnamefont
  {Christensen}}, \bibinfo {author} {\bibfnamefont {C.}~\bibnamefont {Hui}},
  \bibinfo {author} {\bibfnamefont {P.}~\bibnamefont {Jenke}}, \emph {et~al.},\
  }\bibfield  {title} {\bibinfo {title} {Fermi gbm observations of ligo
  gravitational-wave event gw150914},\ }\href@noop {} {\bibfield  {journal}
  {\bibinfo  {journal} {The Astrophysical Journal Letters}\ }\textbf {\bibinfo
  {volume} {826}},\ \bibinfo {pages} {L6} (\bibinfo {year} {2016})}\BibitemShut
  {NoStop}%
\bibitem [{\citenamefont {Greiner}\ \emph {et~al.}(2016)\citenamefont
  {Greiner}, \citenamefont {Burgess}, \citenamefont {Savchenko},\ and\
  \citenamefont {Yu}}]{greiner2016fermi}%
  \BibitemOpen
  \bibfield  {author} {\bibinfo {author} {\bibfnamefont {J.}~\bibnamefont
  {Greiner}}, \bibinfo {author} {\bibfnamefont {J.~M.}\ \bibnamefont
  {Burgess}}, \bibinfo {author} {\bibfnamefont {V.}~\bibnamefont {Savchenko}},\
  and\ \bibinfo {author} {\bibfnamefont {H.-F.}\ \bibnamefont {Yu}},\
  }\bibfield  {title} {\bibinfo {title} {On the fermi-gbm event 0.4 s after
  gw150914},\ }\href@noop {} {\bibfield  {journal} {\bibinfo  {journal} {The
  Astrophysical Journal Letters}\ }\textbf {\bibinfo {volume} {827}},\ \bibinfo
  {pages} {L38} (\bibinfo {year} {2016})}\BibitemShut {NoStop}%
\bibitem [{\citenamefont {Abbott}\ \emph {et~al.}(2016)\citenamefont {Abbott},
  \citenamefont {Abbott}, \citenamefont {Abbott}, \citenamefont {Abernathy},
  \citenamefont {Acernese}, \citenamefont {Ackley}, \citenamefont {Adams},
  \citenamefont {Adams}, \citenamefont {Addesso}, \citenamefont {Adhikari},\
  and\ \citenamefont {et~al.}}]{Abbott_2016}%
  \BibitemOpen
  \bibfield  {author} {\bibinfo {author} {\bibfnamefont {B.}~\bibnamefont
  {Abbott}}, \bibinfo {author} {\bibfnamefont {R.}~\bibnamefont {Abbott}},
  \bibinfo {author} {\bibfnamefont {T.}~\bibnamefont {Abbott}}, \bibinfo
  {author} {\bibfnamefont {M.}~\bibnamefont {Abernathy}}, \bibinfo {author}
  {\bibfnamefont {F.}~\bibnamefont {Acernese}}, \bibinfo {author}
  {\bibfnamefont {K.}~\bibnamefont {Ackley}}, \bibinfo {author} {\bibfnamefont
  {C.}~\bibnamefont {Adams}}, \bibinfo {author} {\bibfnamefont
  {T.}~\bibnamefont {Adams}}, \bibinfo {author} {\bibfnamefont
  {P.}~\bibnamefont {Addesso}}, \bibinfo {author} {\bibfnamefont
  {R.}~\bibnamefont {Adhikari}},\ and\ \bibinfo {author} {\bibnamefont
  {et~al.}},\ }\bibfield  {title} {\bibinfo {title} {Binary black hole mergers
  in the first advanced ligo observing run},\ }\bibfield  {journal} {\bibinfo
  {journal} {Physical Review X}\ }\textbf {\bibinfo {volume} {6}},\ \href
  {https://doi.org/10.1103/physrevx.6.041015} {10.1103/physrevx.6.041015}
  (\bibinfo {year} {2016})\BibitemShut {NoStop}%
\bibitem [{\citenamefont {Meegan}\ \emph {et~al.}(2009)\citenamefont {Meegan},
  \citenamefont {Lichti}, \citenamefont {Bhat}, \citenamefont {Bissaldi},
  \citenamefont {Briggs}, \citenamefont {Connaughton}, \citenamefont {Diehl},
  \citenamefont {Fishman}, \citenamefont {Greiner}, \citenamefont {Hoover}
  \emph {et~al.}}]{meegan2009fermi}%
  \BibitemOpen
  \bibfield  {author} {\bibinfo {author} {\bibfnamefont {C.}~\bibnamefont
  {Meegan}}, \bibinfo {author} {\bibfnamefont {G.}~\bibnamefont {Lichti}},
  \bibinfo {author} {\bibfnamefont {P.}~\bibnamefont {Bhat}}, \bibinfo {author}
  {\bibfnamefont {E.}~\bibnamefont {Bissaldi}}, \bibinfo {author}
  {\bibfnamefont {M.~S.}\ \bibnamefont {Briggs}}, \bibinfo {author}
  {\bibfnamefont {V.}~\bibnamefont {Connaughton}}, \bibinfo {author}
  {\bibfnamefont {R.}~\bibnamefont {Diehl}}, \bibinfo {author} {\bibfnamefont
  {G.}~\bibnamefont {Fishman}}, \bibinfo {author} {\bibfnamefont
  {J.}~\bibnamefont {Greiner}}, \bibinfo {author} {\bibfnamefont {A.~S.}\
  \bibnamefont {Hoover}}, \emph {et~al.},\ }\bibfield  {title} {\bibinfo
  {title} {The fermi gamma-ray burst monitor},\ }\href@noop {} {\bibfield
  {journal} {\bibinfo  {journal} {The Astrophysical Journal}\ }\textbf
  {\bibinfo {volume} {702}},\ \bibinfo {pages} {791} (\bibinfo {year}
  {2009})}\BibitemShut {NoStop}%
\bibitem [{\citenamefont {{Virtanen}}\ \emph {et~al.}(2020)\citenamefont
  {{Virtanen}}, \citenamefont {{Gommers}}, \citenamefont {{Oliphant}},
  \citenamefont {{Haberland}}, \citenamefont {{Reddy}}, \citenamefont
  {{Cournapeau}}, \citenamefont {{Burovski}}, \citenamefont {{Peterson}},
  \citenamefont {{Weckesser}}, \citenamefont {{Bright}}, \citenamefont {{van
  der Walt}}, \citenamefont {{Brett}}, \citenamefont {{Wilson}}, \citenamefont
  {{Jarrod Millman}}, \citenamefont {{Mayorov}}, \citenamefont {{Nelson}},
  \citenamefont {{Jones}}, \citenamefont {{Kern}}, \citenamefont {{Larson}},
  \citenamefont {{Carey}}, \citenamefont {{Polat}}, \citenamefont {{Feng}},
  \citenamefont {{Moore}}, \citenamefont {{Vand erPlas}}, \citenamefont
  {{Laxalde}}, \citenamefont {{Perktold}}, \citenamefont {{Cimrman}},
  \citenamefont {{Henriksen}}, \citenamefont {{Quintero}}, \citenamefont
  {{Harris}}, \citenamefont {{Archibald}}, \citenamefont {{Ribeiro}},
  \citenamefont {{Pedregosa}}, \citenamefont {{van Mulbregt}},\ and\
  \citenamefont {{Contributors}}}]{scipy:2020}%
  \BibitemOpen
  \bibfield  {author} {\bibinfo {author} {\bibfnamefont {P.}~\bibnamefont
  {{Virtanen}}}, \bibinfo {author} {\bibfnamefont {R.}~\bibnamefont
  {{Gommers}}}, \bibinfo {author} {\bibfnamefont {T.~E.}\ \bibnamefont
  {{Oliphant}}}, \bibinfo {author} {\bibfnamefont {M.}~\bibnamefont
  {{Haberland}}}, \bibinfo {author} {\bibfnamefont {T.}~\bibnamefont
  {{Reddy}}}, \bibinfo {author} {\bibfnamefont {D.}~\bibnamefont
  {{Cournapeau}}}, \bibinfo {author} {\bibfnamefont {E.}~\bibnamefont
  {{Burovski}}}, \bibinfo {author} {\bibfnamefont {P.}~\bibnamefont
  {{Peterson}}}, \bibinfo {author} {\bibfnamefont {W.}~\bibnamefont
  {{Weckesser}}}, \bibinfo {author} {\bibfnamefont {J.}~\bibnamefont
  {{Bright}}}, \bibinfo {author} {\bibfnamefont {S.~J.}\ \bibnamefont {{van der
  Walt}}}, \bibinfo {author} {\bibfnamefont {M.}~\bibnamefont {{Brett}}},
  \bibinfo {author} {\bibfnamefont {J.}~\bibnamefont {{Wilson}}}, \bibinfo
  {author} {\bibfnamefont {K.}~\bibnamefont {{Jarrod Millman}}}, \bibinfo
  {author} {\bibfnamefont {N.}~\bibnamefont {{Mayorov}}}, \bibinfo {author}
  {\bibfnamefont {A.~R.~J.}\ \bibnamefont {{Nelson}}}, \bibinfo {author}
  {\bibfnamefont {E.}~\bibnamefont {{Jones}}}, \bibinfo {author} {\bibfnamefont
  {R.}~\bibnamefont {{Kern}}}, \bibinfo {author} {\bibfnamefont
  {E.}~\bibnamefont {{Larson}}}, \bibinfo {author} {\bibfnamefont
  {C.}~\bibnamefont {{Carey}}}, \bibinfo {author} {\bibfnamefont
  {{\.I}.}~\bibnamefont {{Polat}}}, \bibinfo {author} {\bibfnamefont
  {Y.}~\bibnamefont {{Feng}}}, \bibinfo {author} {\bibfnamefont {E.~W.}\
  \bibnamefont {{Moore}}}, \bibinfo {author} {\bibfnamefont {J.}~\bibnamefont
  {{Vand erPlas}}}, \bibinfo {author} {\bibfnamefont {D.}~\bibnamefont
  {{Laxalde}}}, \bibinfo {author} {\bibfnamefont {J.}~\bibnamefont
  {{Perktold}}}, \bibinfo {author} {\bibfnamefont {R.}~\bibnamefont
  {{Cimrman}}}, \bibinfo {author} {\bibfnamefont {I.}~\bibnamefont
  {{Henriksen}}}, \bibinfo {author} {\bibfnamefont {E.~A.}\ \bibnamefont
  {{Quintero}}}, \bibinfo {author} {\bibfnamefont {C.~R.}\ \bibnamefont
  {{Harris}}}, \bibinfo {author} {\bibfnamefont {A.~M.}\ \bibnamefont
  {{Archibald}}}, \bibinfo {author} {\bibfnamefont {A.~H.}\ \bibnamefont
  {{Ribeiro}}}, \bibinfo {author} {\bibfnamefont {F.}~\bibnamefont
  {{Pedregosa}}}, \bibinfo {author} {\bibfnamefont {P.}~\bibnamefont {{van
  Mulbregt}}},\ and\ \bibinfo {author} {\bibfnamefont {S.~.~.}\ \bibnamefont
  {{Contributors}}},\ }\bibfield  {title} {\bibinfo {title} {{SciPy 1.0:
  Fundamental Algorithms for Scientific Computing in Python}},\ }\href
  {https://doi.org/https://doi.org/10.1038/s41592-019-0686-2} {\bibfield
  {journal} {\bibinfo  {journal} {Nature Methods}\ }\textbf {\bibinfo {volume}
  {17}},\ \bibinfo {pages} {261} (\bibinfo {year} {2020})}\BibitemShut
  {NoStop}%
\bibitem [{\citenamefont {Oliphant}(2006)}]{oliphant2006guide}%
  \BibitemOpen
  \bibfield  {author} {\bibinfo {author} {\bibfnamefont {T.~E.}\ \bibnamefont
  {Oliphant}},\ }\href@noop {} {\emph {\bibinfo {title} {{A guide to
  NumPy}}}},\ Vol.~\bibinfo {volume} {1}\ (\bibinfo  {publisher} {Trelgol
  Publishing USA},\ \bibinfo {year} {2006})\BibitemShut {NoStop}%
\bibitem [{\citenamefont {Van Der~Walt}\ \emph {et~al.}(2011)\citenamefont {Van
  Der~Walt}, \citenamefont {Colbert},\ and\ \citenamefont
  {Varoquaux}}]{van2011numpy}%
  \BibitemOpen
  \bibfield  {author} {\bibinfo {author} {\bibfnamefont {S.}~\bibnamefont {Van
  Der~Walt}}, \bibinfo {author} {\bibfnamefont {S.~C.}\ \bibnamefont
  {Colbert}},\ and\ \bibinfo {author} {\bibfnamefont {G.}~\bibnamefont
  {Varoquaux}},\ }\bibfield  {title} {\bibinfo {title} {The numpy array: a
  structure for efficient numerical computation},\ }\href@noop {} {\bibfield
  {journal} {\bibinfo  {journal} {Computing in Science \& Engineering}\
  }\textbf {\bibinfo {volume} {13}},\ \bibinfo {pages} {22} (\bibinfo {year}
  {2011})}\BibitemShut {NoStop}%
\bibitem [{\citenamefont {Harris}\ \emph {et~al.}(2020)\citenamefont {Harris},
  \citenamefont {Millman}, \citenamefont {van~der Walt}, \citenamefont
  {Gommers}, \citenamefont {Virtanen}, \citenamefont {Cournapeau},
  \citenamefont {Wieser}, \citenamefont {Taylor}, \citenamefont {Berg},
  \citenamefont {Smith} \emph {et~al.}}]{harris2020array}%
  \BibitemOpen
  \bibfield  {author} {\bibinfo {author} {\bibfnamefont {C.~R.}\ \bibnamefont
  {Harris}}, \bibinfo {author} {\bibfnamefont {K.~J.}\ \bibnamefont {Millman}},
  \bibinfo {author} {\bibfnamefont {S.~J.}\ \bibnamefont {van~der Walt}},
  \bibinfo {author} {\bibfnamefont {R.}~\bibnamefont {Gommers}}, \bibinfo
  {author} {\bibfnamefont {P.}~\bibnamefont {Virtanen}}, \bibinfo {author}
  {\bibfnamefont {D.}~\bibnamefont {Cournapeau}}, \bibinfo {author}
  {\bibfnamefont {E.}~\bibnamefont {Wieser}}, \bibinfo {author} {\bibfnamefont
  {J.}~\bibnamefont {Taylor}}, \bibinfo {author} {\bibfnamefont
  {S.}~\bibnamefont {Berg}}, \bibinfo {author} {\bibfnamefont {N.~J.}\
  \bibnamefont {Smith}}, \emph {et~al.},\ }\bibfield  {title} {\bibinfo {title}
  {Array programming with numpy},\ }\href@noop {} {\bibfield  {journal}
  {\bibinfo  {journal} {Nature}\ }\textbf {\bibinfo {volume} {585}},\ \bibinfo
  {pages} {357} (\bibinfo {year} {2020})}\BibitemShut {NoStop}%
\bibitem [{\citenamefont {Robitaille}\ \emph {et~al.}(2013)\citenamefont
  {Robitaille}, \citenamefont {Tollerud}, \citenamefont {Greenfield},
  \citenamefont {Droettboom}, \citenamefont {Bray}, \citenamefont {Aldcroft},
  \citenamefont {Davis}, \citenamefont {Ginsburg}, \citenamefont
  {Price-Whelan}, \citenamefont {Kerzendorf} \emph
  {et~al.}}]{robitaille2013astropy}%
  \BibitemOpen
  \bibfield  {author} {\bibinfo {author} {\bibfnamefont {T.~P.}\ \bibnamefont
  {Robitaille}}, \bibinfo {author} {\bibfnamefont {E.~J.}\ \bibnamefont
  {Tollerud}}, \bibinfo {author} {\bibfnamefont {P.}~\bibnamefont
  {Greenfield}}, \bibinfo {author} {\bibfnamefont {M.}~\bibnamefont
  {Droettboom}}, \bibinfo {author} {\bibfnamefont {E.}~\bibnamefont {Bray}},
  \bibinfo {author} {\bibfnamefont {T.}~\bibnamefont {Aldcroft}}, \bibinfo
  {author} {\bibfnamefont {M.}~\bibnamefont {Davis}}, \bibinfo {author}
  {\bibfnamefont {A.}~\bibnamefont {Ginsburg}}, \bibinfo {author}
  {\bibfnamefont {A.~M.}\ \bibnamefont {Price-Whelan}}, \bibinfo {author}
  {\bibfnamefont {W.~E.}\ \bibnamefont {Kerzendorf}}, \emph {et~al.},\
  }\bibfield  {title} {\bibinfo {title} {Astropy: A community python package
  for astronomy},\ }\href@noop {} {\bibfield  {journal} {\bibinfo  {journal}
  {Astronomy \& Astrophysics}\ }\textbf {\bibinfo {volume} {558}},\ \bibinfo
  {pages} {A33} (\bibinfo {year} {2013})}\BibitemShut {NoStop}%
\bibitem [{\citenamefont {Price-Whelan}\ \emph {et~al.}(2018)\citenamefont
  {Price-Whelan}, \citenamefont {Sip{\H{o}}cz}, \citenamefont {G{\"u}nther},
  \citenamefont {Lim}, \citenamefont {Crawford}, \citenamefont {Conseil},
  \citenamefont {Shupe}, \citenamefont {Craig}, \citenamefont {Dencheva},
  \citenamefont {Ginsburg} \emph {et~al.}}]{price2018astropy}%
  \BibitemOpen
  \bibfield  {author} {\bibinfo {author} {\bibfnamefont {A.~M.}\ \bibnamefont
  {Price-Whelan}}, \bibinfo {author} {\bibfnamefont {B.}~\bibnamefont
  {Sip{\H{o}}cz}}, \bibinfo {author} {\bibfnamefont {H.}~\bibnamefont
  {G{\"u}nther}}, \bibinfo {author} {\bibfnamefont {P.}~\bibnamefont {Lim}},
  \bibinfo {author} {\bibfnamefont {S.}~\bibnamefont {Crawford}}, \bibinfo
  {author} {\bibfnamefont {S.}~\bibnamefont {Conseil}}, \bibinfo {author}
  {\bibfnamefont {D.}~\bibnamefont {Shupe}}, \bibinfo {author} {\bibfnamefont
  {M.}~\bibnamefont {Craig}}, \bibinfo {author} {\bibfnamefont
  {N.}~\bibnamefont {Dencheva}}, \bibinfo {author} {\bibfnamefont
  {A.}~\bibnamefont {Ginsburg}}, \emph {et~al.},\ }\bibfield  {title} {\bibinfo
  {title} {The astropy project: Building an open-science project and status of
  the v2. 0 core package},\ }\href@noop {} {\bibfield  {journal} {\bibinfo
  {journal} {The Astronomical Journal}\ }\textbf {\bibinfo {volume} {156}},\
  \bibinfo {pages} {123} (\bibinfo {year} {2018})}\BibitemShut {NoStop}%
\bibitem [{\citenamefont {{Ashton}}\ \emph {et~al.}(2019)\citenamefont
  {{Ashton}}, \citenamefont {{H{\"u}bner}}, \citenamefont {{Lasky}},
  \citenamefont {{Talbot}}, \citenamefont {{Ackley}}, \citenamefont
  {{Biscoveanu}}, \citenamefont {{Chu}}, \citenamefont {{Divakarla}},
  \citenamefont {{Easter}}, \citenamefont {{Goncharov}}, \citenamefont
  {{Hernandez Vivanco}}, \citenamefont {{Harms}}, \citenamefont {{Lower}},
  \citenamefont {{Meadors}}, \citenamefont {{Melchor}}, \citenamefont
  {{Payne}}, \citenamefont {{Pitkin}}, \citenamefont {{Powell}}, \citenamefont
  {{Sarin}}, \citenamefont {{Smith}},\ and\ \citenamefont {{Thrane}}}]{bilby}%
  \BibitemOpen
  \bibfield  {author} {\bibinfo {author} {\bibfnamefont {G.}~\bibnamefont
  {{Ashton}}}, \bibinfo {author} {\bibfnamefont {M.}~\bibnamefont
  {{H{\"u}bner}}}, \bibinfo {author} {\bibfnamefont {P.~D.}\ \bibnamefont
  {{Lasky}}}, \bibinfo {author} {\bibfnamefont {C.}~\bibnamefont {{Talbot}}},
  \bibinfo {author} {\bibfnamefont {K.}~\bibnamefont {{Ackley}}}, \bibinfo
  {author} {\bibfnamefont {S.}~\bibnamefont {{Biscoveanu}}}, \bibinfo {author}
  {\bibfnamefont {Q.}~\bibnamefont {{Chu}}}, \bibinfo {author} {\bibfnamefont
  {A.}~\bibnamefont {{Divakarla}}}, \bibinfo {author} {\bibfnamefont {P.~J.}\
  \bibnamefont {{Easter}}}, \bibinfo {author} {\bibfnamefont {B.}~\bibnamefont
  {{Goncharov}}}, \bibinfo {author} {\bibfnamefont {F.}~\bibnamefont
  {{Hernandez Vivanco}}}, \bibinfo {author} {\bibfnamefont {J.}~\bibnamefont
  {{Harms}}}, \bibinfo {author} {\bibfnamefont {M.~E.}\ \bibnamefont
  {{Lower}}}, \bibinfo {author} {\bibfnamefont {G.~D.}\ \bibnamefont
  {{Meadors}}}, \bibinfo {author} {\bibfnamefont {D.}~\bibnamefont
  {{Melchor}}}, \bibinfo {author} {\bibfnamefont {E.}~\bibnamefont {{Payne}}},
  \bibinfo {author} {\bibfnamefont {M.~D.}\ \bibnamefont {{Pitkin}}}, \bibinfo
  {author} {\bibfnamefont {J.}~\bibnamefont {{Powell}}}, \bibinfo {author}
  {\bibfnamefont {N.}~\bibnamefont {{Sarin}}}, \bibinfo {author} {\bibfnamefont
  {R.~J.~E.}\ \bibnamefont {{Smith}}},\ and\ \bibinfo {author} {\bibfnamefont
  {E.}~\bibnamefont {{Thrane}}},\ }\bibfield  {title} {\bibinfo {title}
  {{BILBY: A User-friendly Bayesian Inference Library for Gravitational-wave
  Astronomy}},\ }\href {https://doi.org/10.3847/1538-4365/ab06fc} {\bibfield
  {journal} {\bibinfo  {journal} {\apjs}\ }\textbf {\bibinfo {volume} {241}},\
  \bibinfo {eid} {27} (\bibinfo {year} {2019})}\BibitemShut {NoStop}%
\end{thebibliography}%

\end{document}